\documentclass{aa}
\usepackage{graphicx}
  \def\kmsec {\rm km~s$^{-1}$}  					
   			 		 
  \def\cmtwo {cm$^{-2}\,$} 			 		 
  \def \al {\rm et al. } 					
  \def \13CO {$^{13}$CO\, }  					
  \def \C18O {C$^{18}$O\, }

  \def \Tmb {$T_{\rm mb}\,$}  					
  
  \def \Trot {$T_{\rm rot}\,$}  
  \def \vlsr {$v_{\rm lsr}\,$}

  \def \Ncol {$N_{\rm col}\,$}

  \def \arcsec {$^{\prime\prime}$\ } 	
  \def \arcmin {$^\prime$\ }  	
  \def \mum {$\mu$m\, }

  \def \3P1 {CI ${\rm ^3P_1-^3P_0}$}

\begin{document}						
 		
\title{A spectral survey of the Orion Nebula from 455 -- 507 GHz}	
 		
\author{Glenn J. White\inst{1,2,3}, M. Araki\inst{4}, J.S. Greaves\inst{5}, M. Ohishi \inst{6},
N.S. Higginbottom\inst{7}} 

\institute{ Centre for Astrophysics \& Planetary Science, University of Kent, Canterbury, Kent
CT2 7NR, England \and Stockholm Observatory, SE--133\,36 Saltsj{\"o}baden, Sweden \and Astrophysics Group, The Cavendish Laboratory, University of Cambridge, Madingley Road, Cambridge CB3 OHE, England \and Institute for Physical Chemistry, University of Basel, Klingelbergstrasse 80, CH-4056 Basel, Switzerland \and Royal Observatory, Blackford Hill, Edinburgh EH9 3HJ, United Kingdom \and National Astronomical Observatory of Japan, 2-21-1, Osawa, Mitaka, Tokyo, 181-8588 Japan \and Department of Physics, Queen Mary \& Westfield College, University of London, Mile End Road, London E1 4NS, England}
 		
\offprints{g.j.white@kent.ac.uk}   		
 		
\date{Received {11th May 2002} / Accepted {22nd May 2003}}   

\titlerunning{Orion spectral survey}
\authorrunning{White et al}	

\abstract{
The results of a submillimetre wavelength spectral line survey between 455.1 -- 507.4 GHz of the Orion-KL hot cloud core are reported. A total of 254 lines were detected to a main beam brightness temperature sensitivity \Tmb $\sim$ 1 - 3 K. The detected lines are identified as being associated with 30 different molecular species or their isotopomeric variants. The strongest line detected was the $J$ = 4--3 transition of the CO molecule. Apart from abundant diatomic rotors such as CO and CS, the spectrum is dominated by SO, SO$_{\rm 2}$ and CH$_{\rm 3}$OH and large organic molecules such as (CH$_3$)$_2$O, CH$_{\rm 3}$CN, C$_{\rm 2}$H$_{\rm 3}$CN, C$_{\rm 2}$H$_{\rm 5}$CN and HCOOCH$_{\rm 3}$ which make up $\sim$ 72\% of the total number of lines; unidentified lines $\sim$ 13\%; and other lines the remaining $\sim$ 15$\%$ of the total. Rotational temperatures and column densities derived using standard rotation diagram analysis techniques were found to range from 70 - 300 K, and 10$^{14}$ -- 10$^{{\rm 17}}$cm$^{\rm 2}$ respectively.
\keywords{Molecules -- star formation -- molecular cloud}
}		
\maketitle
\markboth{Orion submillimetre wavelength spectral survey}{Orion submillimetre wavelength spectral survey}

\section{Introduction}

The chemistry of the Orion-KL molecular cloud core has been better studied than that of any other massive star formation region in the Galaxy [high spectral resolution spectroscopic surveys have been carried out by a number of authors including: 72 - 91 GHz Johansson \al (1984), 70 - 115 GHz Turner (1989),
138 - 151 GHz Lee, Cho, \& Lee (2001), 150 - 160 GHz Ziurys \& McGonagle (1993), 215 - 247 GHz Sutton \al (1985), 216 - 242 GHz Blake \al (1986), 247 - 263 GHz Blake \al (1987), 257 - 273 GHz Greaves \& White (1991), 330 - 360 GHz Jewell \al (1989), 325 - 360 GHz Schilke \al (1997), 342 - 359 GHz White \al (1986),
334 - 343 GHz Sutton \al (1995), 607 - 725 GHz Schilke \al (2001), 780 - 900 GHz Comito \al (2003), 190 - 900 GHz Serabyn \& Weisstein (1995)]. Spectral line surveys can provide an unbiased view of the molecular constituents of the gas in star forming regions, and may be used to estimate the physical and chemical environment. We report here the first high spectral resolution survey in the 600 and 650 $\mu$m atmospheric windows between frequencies of 455 and 507 GHz.

\section{The Data}

The spectral line survey was made using the James Clerk Maxwell telescope in Hawaii during October 1993 over the frequency range 455.1 -- 507.4 GHz. This survey extended across the most of the two atmospheric transmission windows near 650 and 600 $\mu$m. These windows are bracketed by strong telluric H$_2$O absorption lines, and their transparency is highly dependent on the line of sight water vapour column. The data were collected using the JCMT facility receiver, RxC, operated in double-sideband mode. The adopted `on-source' position was that of the `hot core' close to IRc2 ($\alpha$,\,$\delta$)$_{1950}$ = 5$^h$ 32$^m$ 46$^s$.9, -5$^\circ$ 24\arcmin 23$^{\prime\prime}$. The pointing accuracy was measured to be good to better than 2\arcsec rms, from observations of planets and compact calibrator sources used as standards at the telescope.

The half power beam width and main beam efficiencies of the telescope were measured from observations of Mars, Jupiter and Uranus. These ranged from 11\arcsec and 0.53 at the low frequency end of the spectral region to 10\arcsec and 0.49 at the upper end of the band. The receiver double sideband system noise temperatures were typically 1000 - 3000 K. The IF frequency was 3.94 GHz, and the spectra were processed using the JCMT facility 512 MHz bandwidth acousto-optical spectrometer, giving an effective spectral resolution of $\sim$ 0.6 km s$^{-1}$. The spectral region was covered by stepping the local oscillator in 100 MHz steps across the whole spectral region - so that any part of the spectrum was redundantly observed at least four times (i.e. least twice in each sideband - and in many cases more times). Each observation consisted of 4 - 10 minutes integration (total on and off source), which was carried out in a `position-switched' mode, where the telescope was alternated between the on-source position, and a `reference position' located 2100\arcsec to the north. Previous observations of this `reference position' have shown it to be free of significant molecular emission intense enough to affect the accuracy of the survey. The reference position was checked by position switching the telescope against several other positions that were located more than 10 degrees away from the Galactic Plane - and not known to be associated with the locations of any molecular clouds or enhanced interstellar extinction. The spectra were calibrated channel by channel using the standard JCMT three temperature chopper calibration scheme (hot and cold loads and the atmosphere). Observations of the sideband gains were measured by observations of spectral lines that were present in both sidebands. The main beam brightness temperature noise levels varied from 1 - 4 K in a 2 MHz channel ($\sim$ 1.3 \kmsec).

During the data reduction, we attempted to recover an estimate of the single-sideband spectrum, using data collected with a double-sideband receiver. A difficulty common to sideband-deconvolution techniques is the uniqueness of the deconvolution given the observational factors such as pointing reproducibility, sideband gain imbalances, variable calibration solutions as sky conditions change and contamination or blending with strong lines (of both terrestrial and extra-terrestrial origin) in the opposite sideband. A deconvolution technique was used that separated out the emission into the individual sidebands. The basis of this technique (which has been widely used by many observers at the JCMT and forms part of the facility software - although this paper gives the first description of the algorithm that is used) is to set up a series of linked equations for each channel in the DSB spectrum. The first
equation simply describes that the DSB line temperature is the sum of two intensities, one from the upper and one from the lower sideband ($T_{\rm u}$ and $T_{\rm l}$, say). The second equation refers to the same spectral channel but with a shift in the local oscillator setting by $\Delta \nu$: the DSB signal is then the sum of $T_u$ and $T_{\rm (l + 2 \Delta \nu)}$, if we consider the upper sideband to be the frame of reference and the shift to be positive (increased frequency). Similarly we can consider the lower sideband to be the frame of reference, and obtain a third DSB signal that is the sum of $T_{\rm (u + 2 \Delta \nu)} + T_{\rm l}$. This equation set can be extended as far as desired by taking any of the line frequencies offset by $2 \Delta \nu$ and establishing which upper and lower sideband frequencies contribute to the observed DSB signal. The result is always to establish a set of $n$ equations with ($n$ + 1) variables that are linked to a particular channel of the DSB spectrum. A solution can only be found where one of the T(DSB) is consistent with zero within the noise level of the observations. Since this implies (if there are no absorption lines) that both $T_{\rm u}$ and $T_{\rm l}$ are also zero within the noise, the equation set reduces to ($n$ -- 1) unknowns and two known values - hence all $n$ equations can be solved. In the present survey, the local oscillator was stepped in 100 MHz intervals, and the number of equations used per spectral channel was typically 4 - 5. For several spectrally crowded regions, additional spectra were taken with different local oscillator offsets to improve the reliability of the deconvolution.

Multiple coverage of individual parts of the spectrum provided sufficient redundancy to allow
single sideband spectra to be reconstructed. achieving an acceptable solution at most frequencies. The veracity of the technique could also be checked as lines from the lower and upper sidebands move in opposite directions in the DSB spectrum as the local oscillator frequency is stepped (in 100 MHz intervals). The deconvolution technique worked acceptably for more than 98\% of the
whole spectral range covered, but the remaining $\sim$ 2\% could not be solved because there was no signal consistent with zero in the DSB signal-set. A greater number of solutions could be found by extending the number of linked equations, but since the solutions are already of the form of sums and differences of DSB signals, the noise level will be increased if more differencing is involved. It is also inherent in the technique that there are a choice of solutions (for $n$ equations and ($n$ -- 1) unknowns), so we have adopted the minimum SSB results and an initial DSB signal $\leq 2 \sigma$ for the `consistent with zero' DSB criterion. This minimises the level of spikes and should ensure that temperature solutions are underestimated by less than $\approx 3 \sigma$. The locations of the parts of the spectrum for which we did not achieve a good deconvolution, are shown as horizontal bars under the spectrum in Figs \ref{spectra1} - \ref{spectra4}. We also visually inspected at the locations in a spectrum where bright lines from the opposite sideband might have left small residual artifacts (sometimes known as `ghost features' - Schilke \al 1997), as well as inspecting the emission in the opposite sideband to the locations of all of the `U-lines'.
\label{calibsection}

\begin{figure*}
\centering
\includegraphics[width=16cm]{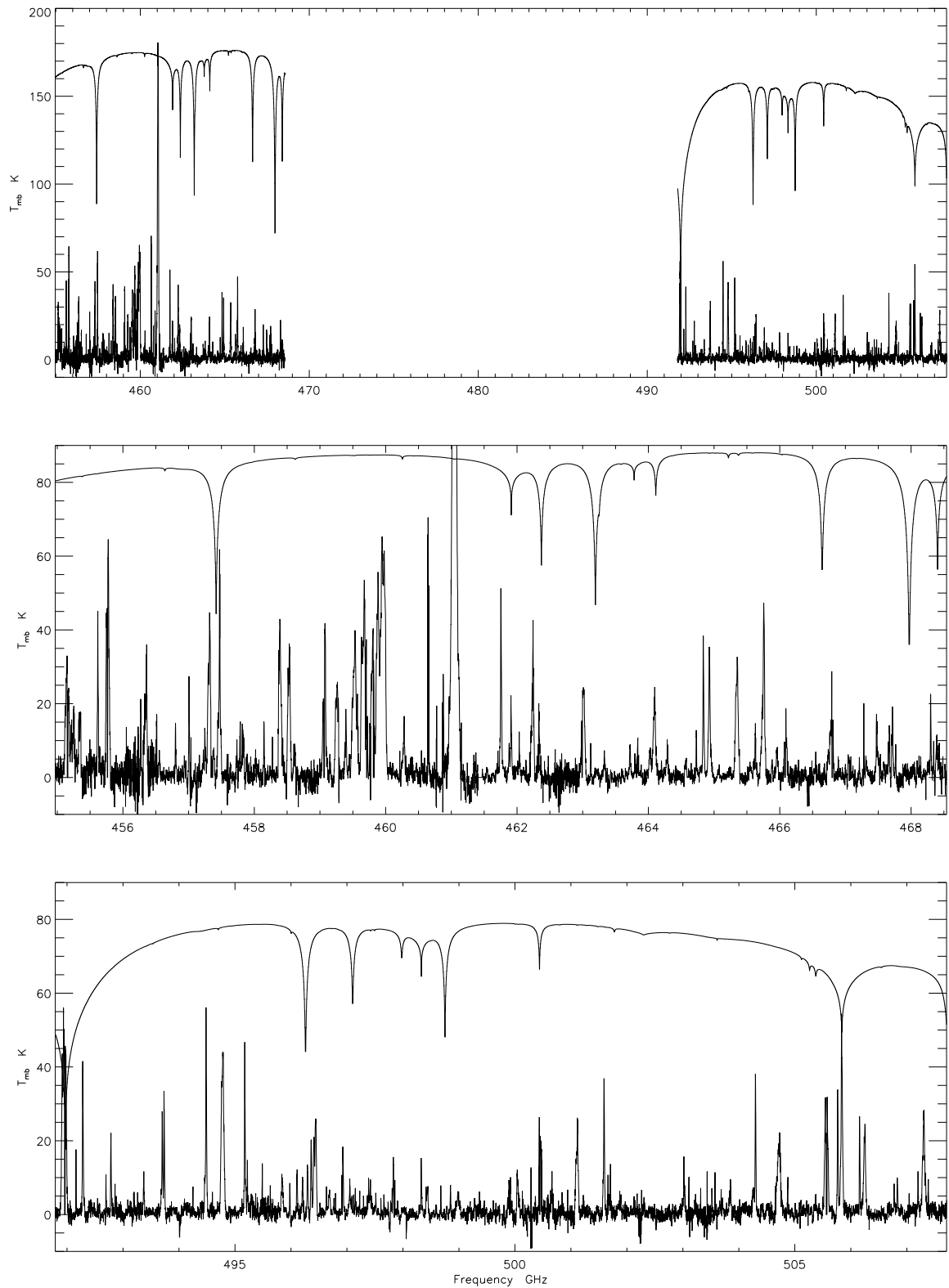}
\caption{The complete spectrum binned in 2 MHz channels in main beam brightness temperature units. A low order polynomial baseline was removed some of the individual spectra. The atmospheric emission spectrum that is overlaid above the Orion spectrum is purely illustrative, and just shows the main features of the atmospheric emission. Recent observational and modeling studies of the terrestrial atmospheric emission (see for example Naylor \al 2000, Pardo, Serabyn \& Cernicharo 2001a) using collisional parameters extracted from the HITRAN database with an independent radiative transfer model (ATM) and different assumptions about line shapes produce broadly similar atmospheric emission spectra to the one shown here.}
\label{composite}
\end{figure*}

\section{Data Analysis}

The intention of this paper is to present the data and some basic results. The spectrum shown
in Fig. \ref{composite} is crowded with many blended lines, and in many
places is confused - reflecting the rich and complex chemistry. A total of 254 lines were identified,
although there may be more lines than this blended together. The lines were clearly identified
using the JPL Sub-millimetre spectral line catalogues of Pickett \al (\cite{pickett:etal}), Pearson \al (1996, 2000), M$\ddot{u}$ller $\al$ (2000) and other lists of line frequencies referred to later in this paper. A total of 98 lines (32\% of
the total) could not be associated with known molecular transitions and have been designated as U-lines.
Although it is possible that some fraction of these may be artifacts of the deconvolution process - the fact that they are distributed throughout the survey, and do not tend to congregate in areas where the
deconvolution process was not successful, it is likely that many of these are real. Table
\ref{linesummary} gives a breakdown of the number of transitions observed from each of the known species.

\begin{table*}
\caption[]{JCMT Orion spectral line survey}
\begin{flushleft}
\begin{tabular}{llllllllll}
\hline\noalign{\smallskip} Molecule & \Ncol & Error & \Trot & Error & Number & Note \\
\noalign{\smallskip} & cm$^{-3}$ & cm$^{-3}$ & K & K & of lines & \\
\noalign{\smallskip}
\hline\noalign{\smallskip}
\noalign{\smallskip}
(CH$_3$)$_2$O & 1.4 10$^{16}$ & 1.8 10$^{15}$ & 157 & 30 & 27(26) lines & \\
C$_2$H$_3$CN & 3.0 10$^{17}$ & 3.2 10$^{17}$ & 180 & 47 & 13(6) lines & [1] \\
C$_2$H$_5$CN & 2.4 10$^{16}$ & 8.4 10$^{15}$ & 150 & 12 & 27(23) lines & T$_{rot}$ from Sut85\\
 & 8.3 10$^{15}$ & 1.2 10$^{15}$ & 239 & 12 & 27(23) lines & T$_{rot}$ from Sch01\\
C$_2$H$_5$OH & 5.6 10$^{16}$ & 4.0 10$^{16}$ & 70 & - & 8(6) lines & T$_{rot}$ from Ohi95 \\
& 4.1 10$^{16}$ & 2.3 10$^{16}$ & 264 & 196 & 8(6) lines & \\
CH$_2$NH & 2.4 10$^{15}$ & 5.9 10$^{14}$ & 150 & - & 3(2) lines & T$_{rot}$ from HNCO \\
CH$_3$CN & 3.6 10$^{15}$ & 4.2 10$^{14}$ & 227 & 21 & 19 lines & \\
CH$_3$$^{13}$CN & 6.0 10$^{14}$ & 2.2 10$^{14}$ & 227 & - & 2 lines & T$_{rot}$ From CH$_3$CN \\
& 2.9 10$^{15}$ & - & 74 & - & 2 lines & \\
CH$_3$OH & 9.3 10$^{16}$ & 4.8 10$^{16}$ & 599 & 295 & 24(23) lines & \\
H$_2$CO & 1.6 10$^{16}$ & - & 166 & - & 2(1) lines & T$_{rot}$ from Bla87 \\
H$_2$$^{13}$CO & 1.0 10$^{15}$ & 4.0 10$^{14}$ & 166 & - & 2 lines & T$_{rot}$ from Bla87 \\
HC$_3$N & 1.5 10$^{15}$ & - & 164 & - & 2 lines & \\
HCOOCH$_3$ & 5.1 10$^{16}$ & 9.5 10$^{15}$ & 301 & 95 & 26(24) lines & \\
HNCO & 4.9 10$^{15}$ & 4.0 10$^{14}$ & 150 & 14 & 4(3) lines & \\
NH$_2$CN & 3.3 10$^{15}$ & 8.5 10$^{14}$ & 200 & - & 3 lines & T$_{rot}$ = 200 (K) Fix \\
& 1.1 10$^{16}$ & 4.8 10$^{15}$ & 100 & - & 3 lines & T$_{rot}$ = 100 (K) Fix \\
OCS & 9.0 10$^{16}$ & - & 106 & - & 2 lines & \\
SO & 3.3 10$^{17}$ & - & 72 & - & 2(1) lines & T$_{rot}$ from Sut95 \\
$^{34}$SO & 1.1 10$^{16}$ & 3.5 10$^{15}$ & 89 & 43 & 5(4) lines & \\
SO$_2$ & 1.2 10$^{17}$ & 1.0 10$^{16}$ & 136 & 9 & 35(28) lines & \\
$^{34}$SO$_2$ & 8.5 10$^{15}$ & - & 156 & - & 3(2) lines & \\
$^{13}$CS & 2.3 10$^{14}$ & - & 120 & - & 1 line & T$_{rot}$ from Zen95 \\
$^{30}$SiO & 3.4 10$^{14}$ & - & 50 & - & 1 line & T$_{rot}$ from Sut95 \\
CH$_3$CHO & 1.1 10$^{16}$ & - & 81 & - & 1 line & T$_{rot}$ from Sch97 \\
CI & 1.2 10$^{18}$ & - & 30 & - & 1 line & T$_{ex}$ from Whi95 \\
CO & 3.5 10$^{18}$ & - & 200 & - & 4 lines (1 transition) & T$_{rot}$ from Sut95 \\
DCN & 1.1 10$^{14}$ & - & 200 & - & 1 line & T$_{rot}$ from Bla87 \\
HCOOH & 2.2 10$^{15}$ & - & 100 & - & 1 line & T$_{rot}$ from Sut95 \\
HDO & 3.2 10$^{16}$ & - & 164 & - & 1 line & T$_{rot}$ from Bla87 \\
N$_2$O & 4.6 10$^{16}$ & - & 230 & - & 1 line & T$_{rot}$ from Wri83 \\
NH$_2$CHO & 7.5 10$^{15}$ & - & 81 & - & 1 lines & T$_{rot}$ from Sch97 \\
NH$_2$D & 8.7 10$^{15}$ & - & 160 & - & 1 line & T$_{rot}$ from Her88 \\
\hline\noalign{\smallskip}
\end{tabular}
\end{flushleft}
LTE rotation temperatures and beam averaged \Ncol, estimated using a Boltzmann plot. The \Ncol were determined using the main beam brightness temperature scale, and for species where only one line was measured, we have assumed \Trot as described in Column 7. Notes: Column 6: A(B) lines represent A: number of assigned lines, B: number of lines included in the fitting. For A lines only, all lines were included in the fitting. [1] : $v_{\rm 2}$ = 1 state lines were not included in the fitting, [2] : Energy levels were considered in the $v$ = 1 vibrational state.	Errors quoted are all 1$\sigma$.
\label{linesummary}
\end{table*} 

These data were used
to estimate the rotational temperature, \Trot, and column densities, \Ncol, of the various
species using the relationship given in Eq. \ref{colden}.

\begin{equation}
ln\left(\frac{3kc\int{T_{mb}}dV}{8{\pi^{3}}{{\mu^{2}}{\nu^{2}}S}}\right)
=ln\left(\frac{{N_{T}}}{Q\left({T_{rot}}\right)}\right)-\frac{{E{_u}}}{k{T_{rot}}}
\label{colden}
\end{equation}

where $\int{T_{\rm mb}} dV$ is the integrated intensity of the line, S is the intrinsic strength
of the transition, Q is the partition function of the molecule and $\mu$ is the dipole moment.
The value of ${\mu^{2}}S$ was calculated from the value of the transition intensity 
listed by Pickett {\rm et al} (1995), using Eq. \ref{intensity}.
\begin{equation}
{I_{ba}}\left(T\right)=\left(\frac{8{\pi^{3}}}{3hc}\right){\nu_{ba}}{^{x}S_{ba}}
{\mu_{x}^{2}}\left[\frac{{e^{\frac{{E_{l}}}{kT}}}-{e^{\frac{{E_{u}}}{kT}}}}{{Q_{rs}}}\right]
\label{intensity}
\end{equation}

The values for \Trot and column densities are summarised in Table \ref{linesummary}.

One of the objectives in analysing spectral line survey data is to determine molecular parameters, such as rotation temperatures, column densities etc. The approach commonly adopted has been to use a `rotation diagram' to estimate these parameters. In this, an optically thin transition produces an antenna temperature that is proportional to the column density in the upper level of the transition being observed. If all transitions are thermalised and the kinetic temperature is known, then a single
integrated line intensity and be used to estimate the total column density of the species in question. The rotational temperature diagram is a plot showing the column density per statistical weight of a number of molecular energy levels, as a function of energy above the ground state - in local thermodynamic equilibrium (LTE), this is equivalent to a Boltzmann distribution. A plot of the natural logarithm of the column density, $N$, divided by the degeneracy $g$, versus the Energy $E$ of the final state of the level (expressed in units of degrees K) $E$/$k$,  will lead to a straight line fit with a slope of 1/$T$, where $g$ is the statistical weight of level $u$ lying at $E$ energy above the ground state, and $T$ is the rotational temperature. This is equivalent to the kinetic
temperature in the limit where all of the levels are thermalised.

One problem for the rotation diagram method is that it may underestimate the total column density if
some of the lines fitted are optically thick, or LTE conditions do not hold, or if the background radiation is non-negligible (Turner 1991, Goldsmith \& Langer 1999, Nummelin \al 2000, Schilke \al 2001).
This can be however be addressed by using less abundant isotopomeric variants that allow estimates to be made of the optical depths, and then using these to correct the column density estimates. It has however been widely used in past molecular line survey studies, because of its computational simplicity, and the absence of a need to have observations of an isotopomer. In this paper we give results from the traditional rotation-diagram technique (Table 1), except in cases pointed out in the Table and text where the results may be affected by high optical depth effects. We have examined a number of cases (CH$_{\rm 3}$CN, SO$_{\rm 2}$, OCS, SO, CH$_{\rm 3}$OH and $^{\rm 34}$SO in some detail - see for example Sections \ref{ch3cnsection} and \ref{so2section} - calculating the optically thick and thin total column densities - and found the use of the rotational temperature technique to provide similar column density estimates are quite similar to the rotational diagram values (see for example discussion of opacities in Section \ref{so2section}). Another indicator as to whether opacity corrections are important, is to look at the rotational temperature diagrams (see Fig \ref{rotemp}), to see whether there are data points that deviate noticeably from a straight line fit - and which may indicate that the line has saturated. Lee, Cho and Lee (2001) applied this kind of constraint to seven molecules (CH$_{\rm 3}$OH, HCOOCH$_{\rm 3}$, CH$_{\rm 3}$OCH$_{\rm 3}$, C$_{\rm 2}$H$_{\rm 5}$CN, SO$_{\rm 2}$, CH$_{\rm 3}$CN, and H$_{\rm 2}$CO), finding that the majority could to first order be treated using the rotational technique. The issue of opacity will be further discussed in the relevant sections dealing with molecules that may be opaque - where we also conclude that the effects of opacity are minor in the analysis of data from this survey, and that our justification of using the optically thin rotational temperature technique is adequate although there clear examples (Schilke \al 1997, 2001) where opacity corrections are required for CH$_{\rm 3}$OH.

\begin{figure}
\resizebox{\hsize}{!}{\includegraphics{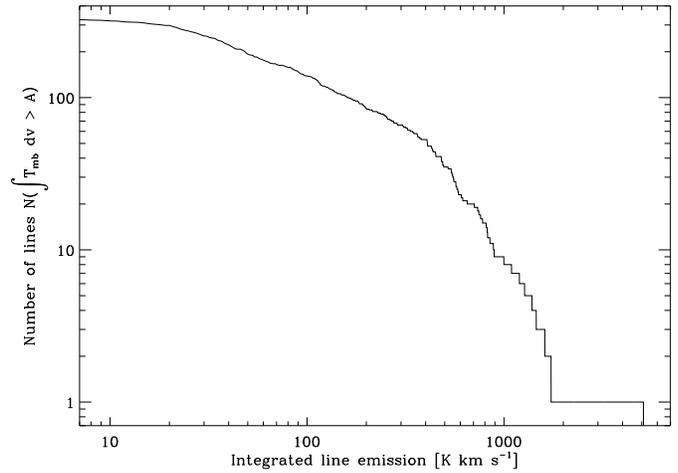}}
\caption{Cumulative spectrum showing the number of lines detected that exceeded some value.}
\label{fullspectrum}
\end{figure}

Full details of the detected lines are given in Tables \ref{allines1} -- \ref{allines5}.

\section{Discussion of Individual Species}

\begin{figure*}
\centering
\includegraphics[width=17cm]{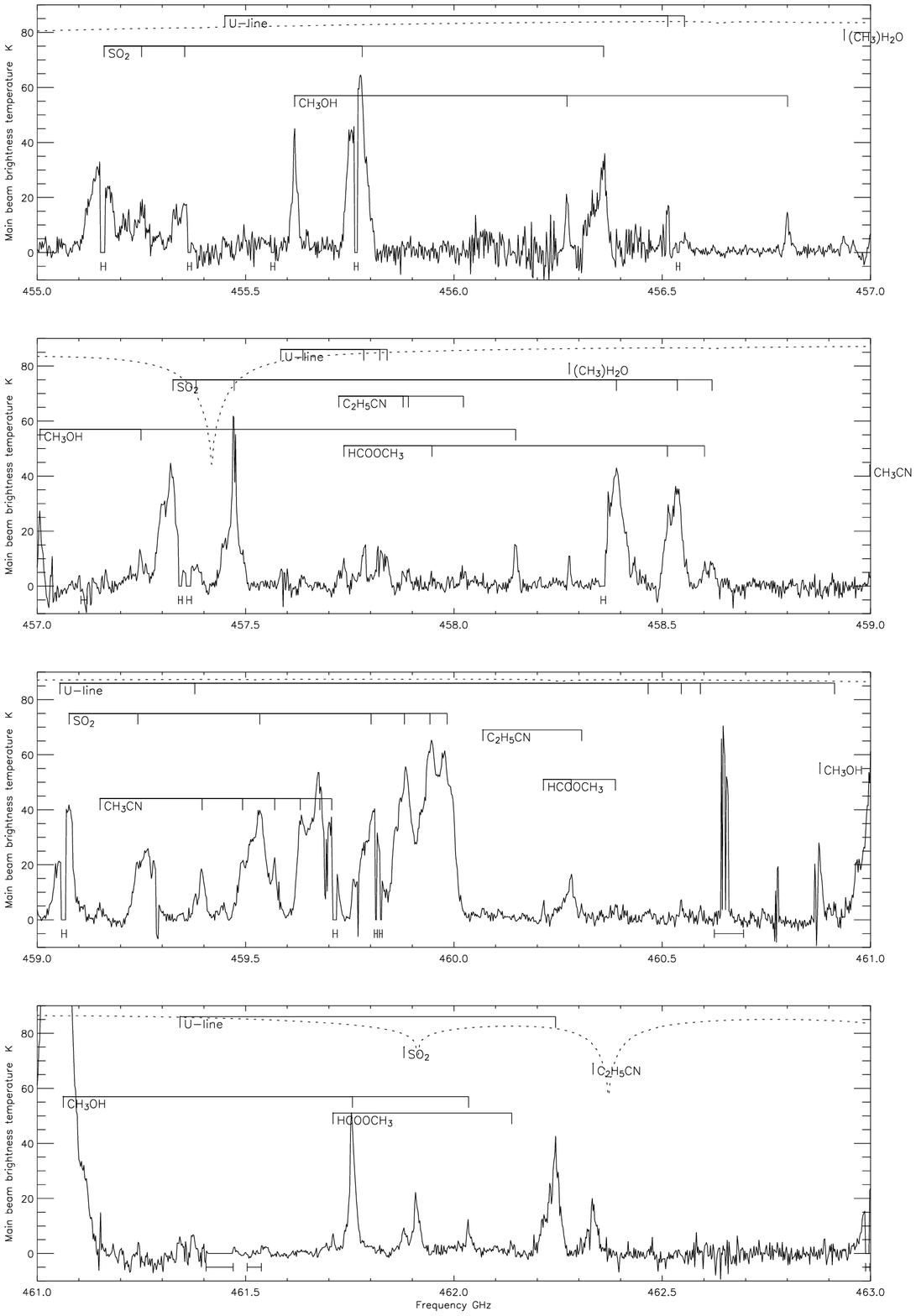}
\caption{Spectra and lines detected in the survey. The locations of parts of the spectra that were poorly deconvolved are indicated by horizontal lines placed underneath the spectrum.}
\label{spectra1}
\end{figure*}

\begin{figure*}
\centering
\includegraphics[width=17cm]{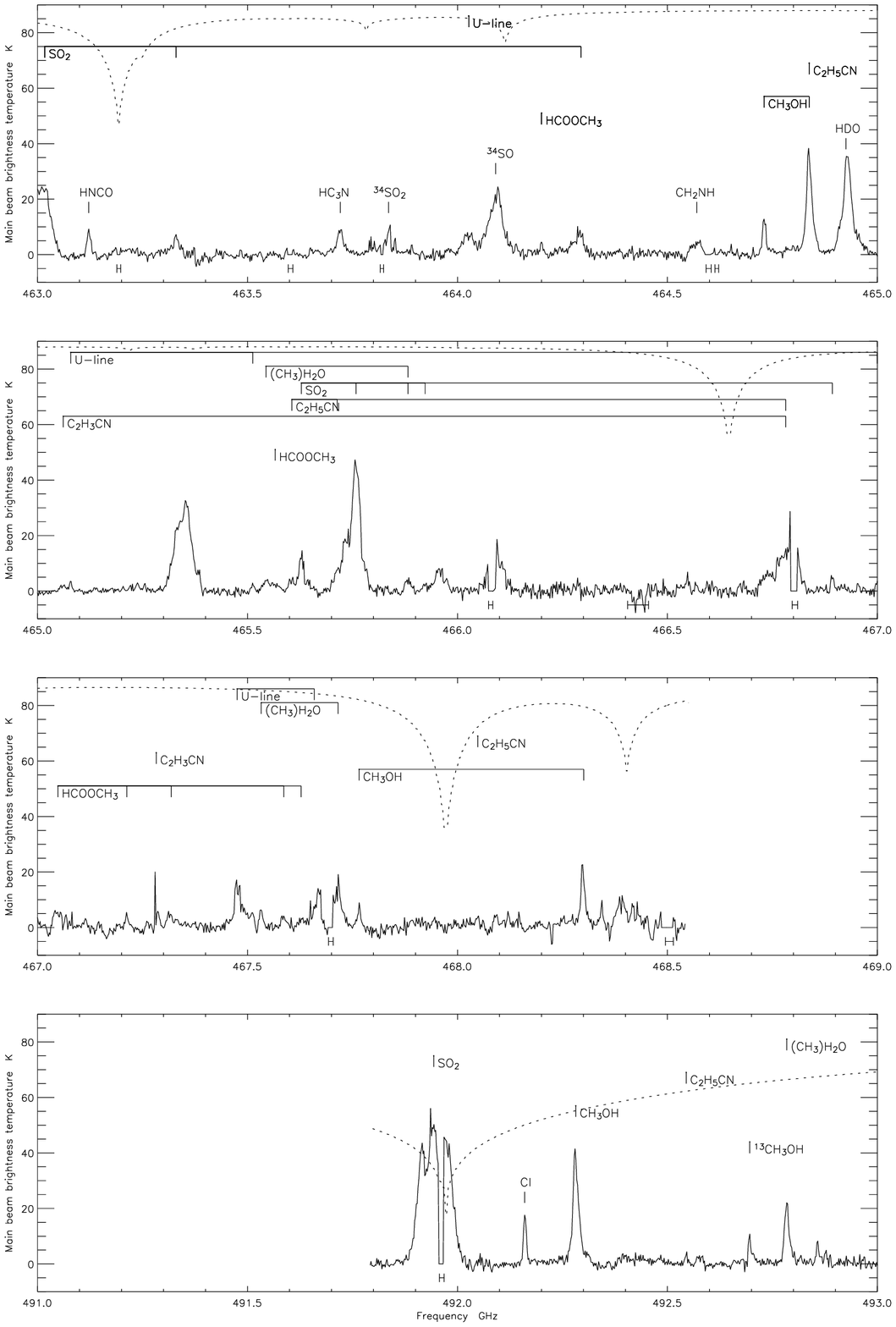}
\caption{Spectra and lines detected in the survey. The locations of parts of the spectra that were poorly deconvolved are indicated by horizontal lines placed underneath the spectrum.}
\label{spectra2}
\end{figure*}

\begin{figure*}
\centering
\includegraphics[width=17cm]{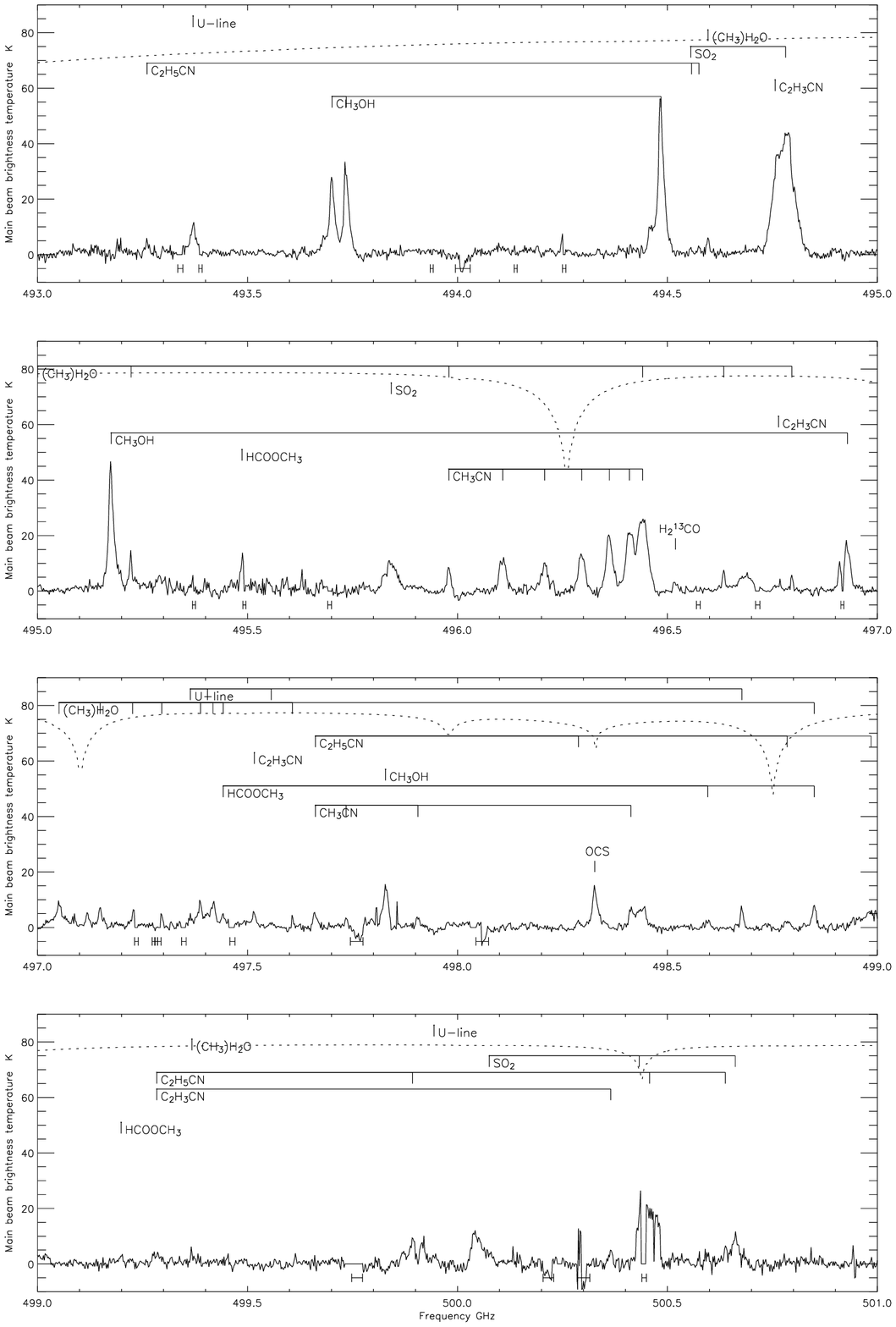}
\caption{Spectra and lines detected in the survey. The locations of parts of the spectra that were poorly deconvolved are indicated by horizontal lines placed underneath the spectrum.}
\label{spectra3}
\end{figure*}

\begin{figure*}
\centering
\includegraphics[width=17cm]{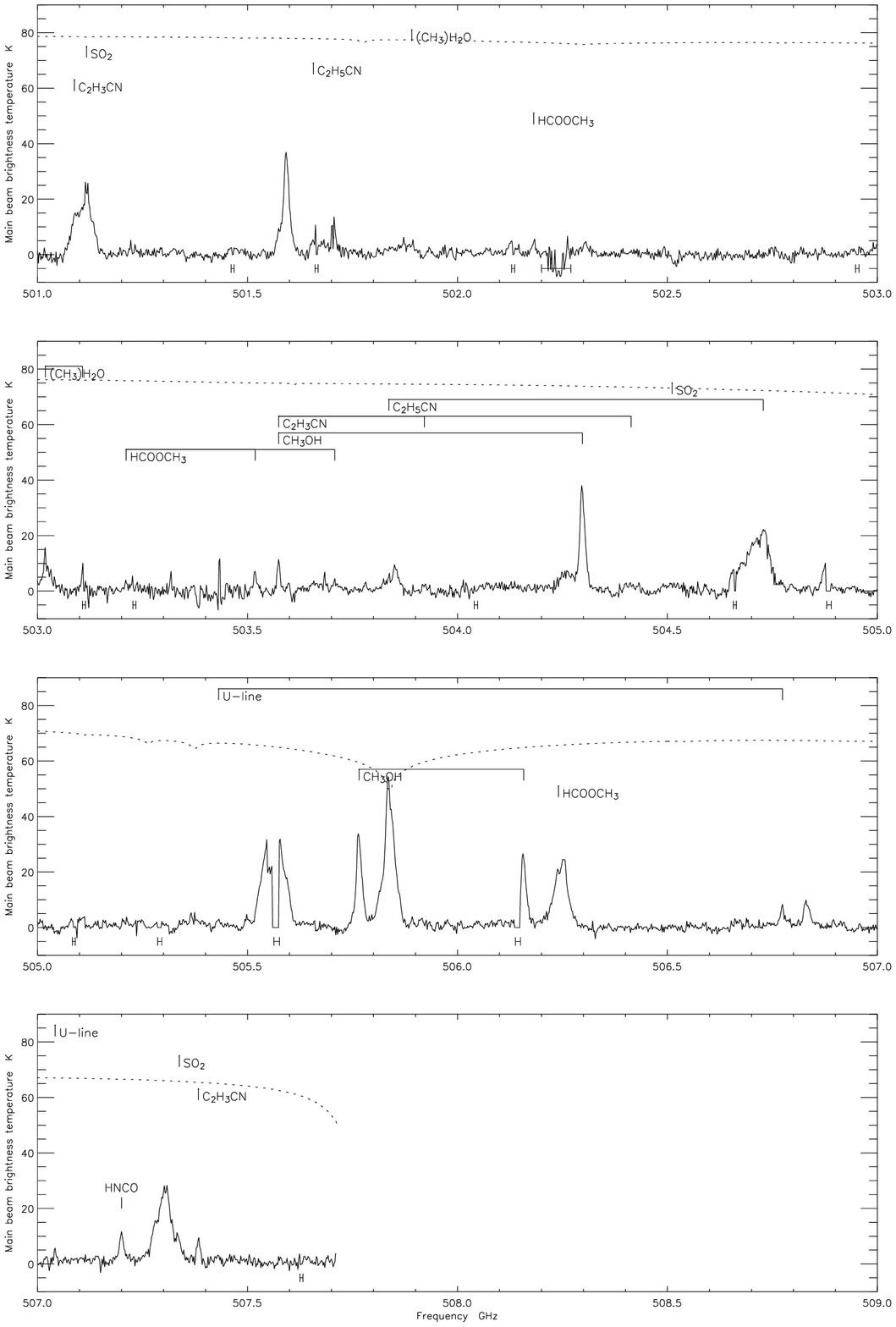}
\caption{Spectra and lines detected in the survey. The locations of parts of the spectra that were poorly deconvolved are indicated by horizontal lines placed underneath the spectrum.}
\label{spectra4}
\end{figure*}

Individual spectra are shown in more detail in Figs. \ref{spectra1}, \ref{spectra2},
\ref{spectra3} and \ref{spectra4}, along with identifications of the prominent lines. Previous studies (Blake {\rm et al. } 1987, Schilke \al 1997, 2001) have shown that there are four characteristic velocity components in Orion spectra: the extended ridge - which is ambient gas in the Orion Molecular cloud (\vlsr $\sim$ 9 \kmsec $\Delta\nu \sim$ 4 \kmsec), the compact ridge - which is a compact clump lying about 10\arcsec south-west of the hot core (\vlsr $\sim$ 8 \kmsec $\Delta\nu \sim$ 3 \kmsec), the plateau - which which has been identified as the outflow, and is associated with the broadest spectral lines (\vlsr $\sim$ 6--10 \kmsec $\Delta\nu \sim$ 20 \kmsec), and the hot core that lies close to the infrared source IRc2 (\vlsr $\sim$ 3--6 \kmsec $\Delta\nu \sim$  5--10 \kmsec). Most of the observed lines show velocities within these ranges, except a few lines which may suffer from blending (e.g. lines at 455.7798, 459.7069, 466.7817, 495.0007, 496.44095, 503.8517, 504.7281 GHz). The superposed dotted line shows the normalised atmospheric absorption (see Fig \ref{composite} caption)
under conditions typical for the survey.

\begin{table*}
\caption[]{JCMT Orion spectral linesurvey}
\begin{flushleft}
\begin{tabular}{lllllllll}
	\hline Frequency&Species&Transition&Peak&Width&Vel&Notes\\
	\noalign{\smallskip} MHz& & & T$_{mb}$ K& km s$^{-1}$& km s$^{-1}$& \\
	\noalign{\smallskip}
	\hline\noalign{\smallskip}
	\noalign{\smallskip}
455.160796 & SO$_2$ & 21$_{5,17}$-21$_{4,18}$ & 23.2 & 33.3 & 8.2 & \\
455.250265 & SO$_2$ & 29$_{5,25}$-29$_{4,26}$ & 14.8 & 19.6 & 6.9 & \\
455.353721 & SO$_2$ & 18$_{5,13}$-18$_{4,14}$ & 17.6 & 8.0 & 5.4 & \\
455.450207 & U-line & & 4.3 & 9.4 & & \\
455.617930 & CH$_3$OH & 6$_{1}$-5$_{0}$ & 36.1 & 10.6 & 9.1 & \\
455.779855 & SO$_2$ & 11$_{3,9}$-10$_{2,8}$ & 60.2 & 13.8 & 1.6 & Blend with C$_2$H$_3$CN\\
456.271440 & CH$_3$OH & 18$_{1}$-17$_{2}$ & 20.2 & 6.2 & 8.4 & \\
456.359253 & SO$_2$ & 19$_{5,15}$-19$_{4,16}$ & 29.5 & 11.6 & 4.4 & \\
456.513228 & U-line & & 18.8 & 4.3 & & \\
456.553270 & U-line & & 4.7 & 16.6 & & C$_3$H$_2$ 22$_{0,22}$-21$_{0,21}$ ?\\
456.800826 & CH$_3$OH & 18$_{-2}$-18$_{-1}$ & 12.7 & 7.9 & 5.3 & Blend with C$_2$H$_5$CN\\
456.936702 & (CH$_3$)$_2$O & 15$_{3,12}$-14$_{2,13}$ & 5.0 & 9.3 & 6.9 & 4 lines blended\\
457.006904 & CH$_3$OH & 11$_{2}$-11$_{1}$ & 23.7 & 7.6 & 8.2 & \\
457.248931 & CH$_3$OH & 19$_{-1}$-18$_{-2}$ & 9.3 & 15.6 & 9.9 & \\
457.325926 & SO$_2$ & 16$_{5,11}$-16$_{4,12}$ & 30.4 & 11.3 & 2.2 & Blend with C$_2$H$_3$CN\\
457.381156 & SO$_2$ & 15$_{8,8}$-16$_{7,9}$+15$_{8,7}$-16$_{7,10}$ & 7.5 & 14.6 & 7.8 & \\
457.472642 & SO$_2$ & 17$_{5,13}$-17$_{4,14}$ & 50.1 & 11.4 & 5.2 & \\
457.584559 & U-line & & 7.0 & 3.5 & & \\
457.637840 & U-line & & 3.5 & 7.3 & & \\
457.723839 & C$_2$H$_5$CN & 51$_{8,44}$-50$_{8,43}$ & 5.1 & 5.2 & 11.5 & \\
457.736126 & HCOOCH$_3$ & 40$_{3,37}$-39$_{3,36}$ & 9.2 & 6.0 & 7.4 & \\
457.783725 & U-line & & 13.3 & 9.4 & & \\
457.822031 & U-line & & 9.1 & 28.5 & & \\
457.839618 & U-line & & 4.7 & 4.3 & & SO$_2$ $v_2$ = 1 ?\\
457.878236 & C$_2$H$_5$CN & 53$_{2,52}$-52$_{2,51}$ & 5.4 & 4.3 & 5.2 & \\
457.890562 & C$_2$H$_5$CN & 28$_{5,23}$-27$_{4,24}$ & 5.9 & 6.0 & 9.3 & \\
457.947140 & HCOOCH$_3$ & 38$_{6,32}$-37$_{5,32}$ & 3.4 & 6.0 & 10.5 &\\
457.985674 & HCOOH & 20$_{3,17}$-19$_{3,16}$ & 3.8 & 2.8 & 9.6 & \\
458.022659 & C$_2$H$_5$CN & 48$_{4,45}$-47$_{3,44}$ & 4.9 & 10.3 & 7.2 & \\
458.148139 & CH$_3$OH & 10$_{1}$-9$_{2}$ & 15.3 & 7.0 & 7.0 & \\
458.276916 & (CH$_3$)$_2$O & 25$_{1,24}$-24$_{2,23}$ & 11.6 & 3.9 & 8.8 & 4 lines blended\\
458.389953 & SO$_2$ & 15$_{5,11}$-15$_{4,12}$ & 40.8 & 25.2 & 6.9 & \\
458.433457 & C$_2$H$_5$OH & 28$_{1,28}$-- 27$_{1,27}$ & 8.2 & 5.3 & 6.1 & \\
458.512326 & HCOOCH$_3$ & 41$_{2/3/2/2,39}$-40$_{3/3/2/2,38}$ & 20.7 & 12.2 & 8.5 & 4 lines blended\\
458.536360 & SO$_2$ & 14$_{5,9}$-14$_{4,10}$ & 34.8 & 16.7 & 5.9 & \\
458.601426 & HCOOCH$_3$ & 36$_{11,25}$-35$_{11,24}$ & 7.1 & 8.0 & 9.9 & \\
458.619675 & SO$_2$ & 31$_{5,27}$-31$_{4,28}$ & 8.6 & 9.8 & 3.8 & \\
458.998198 & CH$_3$CN & $J$=25-24 $K$ = 9 & 3.9 & 5.7 & 7.3 & \\
459.054419 & U-line & & 20.5 & 22.1 & & \\
459.076694 & SO$_2$ & 13$_{5,9}$-13$_{4,10}$ & 36.0 & 14.1 & 5.0 & \\
459.150830 & CH$_3$CN & $J$ = 25-24 $K$ = 8 & 4.3 & 11.2 & 4.9 & \\
459.257959 & U-line & & 23.3 & 24.2 & & \\
459.378018 & U-line & & 8.3 & 6.1 & & \\
459.395715 & CH$_3$CN & $J$=25-24 $K$ = 6 & 17.7 & 10.4 & 5.4 & Blend with CH$_3$$^{13}$CN $J$=25-24 $K$ =3\\
459.443401 & CH$_3$$^{13}$CN & $J$=25-24 $K$ = 2 & 4.7 & 13.9 & 9.6 & \\
459.492742 & CH$_3$CN & $J$=25-24 $K$ =5 & 20.4 & 16.7 & 5.3 & Blend with CH$_3$$^{13}$CN $J$=25-24 $K$ = 0,1\\
459.511802 & NH$_2$CN & 23$_{4,20/19}$-22$_{4,19/18}$ & 16.0 & 9.3 & 9.0 & \\
459.534263 & SO$_2$ & 11$_{5,7}$-11$_{4,8}$ & 39.4 & 19.7 & 5.2 & \\
459.569872 & CH$_3$CN & $J$=25-24 $K$ =4 & 17.3 & 13.6 & 6.6 & \\
459.631405 & CH$_3$CN & $J$=25-24 $K$ =3 & 30.9 & 13.0 & 6.6 & \\
459.678274 & CH$_3$CN & $J$=25-24 $K$ = 2 & 37.8 & 10.8 & 4.8 & \\
459.706888 & CH$_3$CN & $J$=25-24 $K$ = 0,1 & 35.4 & 15.5 & 3.3 & 2 lines blended\\
459.760261 & HNCO & 21$_{1,21}$-20$_{1,20}$ & 14.8 & 7.2 & 6.1 & \\
459.801217 & SO$_2$ & 9$_{5,4}$-9$_{4,5}$+9$_{5,5}$-9$_{4,6}$ & 35.0 & 22.8 & 8.1 & \\
459.881164 & SO$_2$ & 8$_{5,3}$-8$_{4,4}$+8$_{5,4}$-8$_{4,5}$ & 49.9 & 32.7 & 7.6 & \\
459.942528 & SO$_2$ & 7$_{5,3}$-7$_{4,4}$ & 57.9 & 26.2 & 4.8 & \\
459.983548 & SO$_2$ & 5$_{5,0}$-5$_{4,1}$+5$_{5,1}$-5$_{4,2}$ & 53.6 & 25.6 & 11.4 & \\
\hline \noalign{\smallskip}
\label{allines1}
\end{tabular}
\end{flushleft}
\end{table*}

\begin{table*}
\caption[]{JCMT Orion spectral linesurvey}
\begin{flushleft}
\begin{tabular}{lllllllll}
	\hline Frequency&Species&Transition&Peak&Width&Vel&Notes\\
	\noalign{\smallskip} MHz& & & T$_{mb}$ K& km s$^{-1}$& km s$^{-1}$& \\
	\noalign{\smallskip}
	\hline\noalign{\smallskip}
	\noalign{\smallskip}
460.069658 & C$_2$H$_5$CN & 31$_{4,27}$-30$_{3,28}$ & 2.6 & 23.5 & 7.3 & \\
460.214973 & HCOOCH$_3$ & 37$_{6,31}$-36$_{6,30}$ & 6.5 & 3.6 & 8.8 & \\
460.272941 & $^{34}$SO$_2$ & 6$_{4,2}$-5$_{3,3}$ & 7.5 & 31.1 & 7.8 & \\
460.281266 & HCOOCH$_3$ & 43$_{1,43}$-42$_{1,42}$ & 8.8 & 6.9 & 9.6 & \\
460.306531 & C$_2$H$_5$CN & 22$_{15,7/8}$-23$_{14,10/9}$ & 4.0 & 2.3 & 7.1 & \\
460.387575 & HCOOCH$_3$ & 37$_{12,26}$-36$_{12,25}$ & 3.1 & 27.3 & 8.0 & 2 lines blended\\
460.466108 & U-line & & 3.0 & 13.7 & & \\
460.545686 & U-line & & 6.2 & 6.2 & & \\
460.591671 & U-line & & 5.0 & 3.9 & & \\
460.878573 & CH$_3$OH & 9$_{2}$-8$_{1}$ & 24.4 & 6.6 & 7.3 & \\
460.913504 & U-line & & 3.8 & 5.0 & & \\
461.040768 & CO & $J$ = 4-3 & 182.0 & 51.0 & 9.0 & \\
461.342825 & U-line & & 5.0 & 7.5 & & \\
461.373909 & HNCO & 21$_{2,19}$-20$_{2,18}$ & 7.4 & 6.1 & 5.7 & \\
461.709888 & HCOOCH$_3$ & 30$_{7,24}$-29$_{6,23}$ & 5.5 & 8.4 & 9.6 & \\
461.756285 & CH$_3$OH & 15$_{0}$-14$_{1}$& 43.8 & 11.8 & 7.8 & \\
461.880198 & SO$_2$ & 20$_{9,11}$-21$_{8,14}$+20$_{9,12}$-21$_{8,13}$ & 7.9 & 12.6 & 6.2 & \\
461.910483 & OCS & $J$=38-37 & 18.1 & 11.8 & 7.2 & \\
462.035190 & CH$_3$OH $v_t$ = 1 & 10$_{6}$-11$_{5}$+& 8.1 & 13.0 & 9.8 & \\
462.036358 & C$_2$H$_5$OH & 13$_{5,9}$-12$_{4,8}$ & 8.1 & 13.0 & 12.5 & Blended with CH$_3$NC\\
462.138660 & HCOOCH$_3$ & 37$_{11,27}$-36$_{11,26}$ & 3.5 & 7.0 & 8.4 & \\
462.236037 & $^{34}$SO & & 42.0 & 19.3 & & \\
462.334032 & $^{13}$CS & $J$=10-9 & 14.6 & 16.4 & 6.8 & Blended with C$_2$H$_5$CN\\
463.017334 & SO$_2$ & 12$_{2,10}$-11$_{1,11}$ & 21.3 & 23.0 & 5.2 & \\
463.122260 & HNCO & 21$_{1,20}$-20$_{1,19}$ & 8.5 & 6.8 & 7.2 & \\
463.329490 & SO$_2$ & 35$_{4,32}$-35$_{3,33}$ & 5.6 & 14.3 & 6.9 & \\
463.720861 & HC$_3$N & $J$=51-50 & 8.2 & 13.2 & 5.4 & \\
463.836119 & $^{34}$SO$_2$ & 26$_{0,26}$-25$_{1,25}$ & 8.2 & 10.4 & 9.1 & \\
464.026804 & U-line & & 6.8 & 25.2 & & \\
464.091929 & $^{34}$SO & N$_{J}$=11$_{11}$-10$_{10}$ & 20.8 & 25.8 & 11.6 & \\
464.200231 & CH$_3$CHO & 7$_{4,4}$-6$_{3,3}$ & 4.2 & 5.8 & 7.9 & Blend with HCOOCH$_3$\\
464.200231 & HCOOCH$_3$ & 37$_{10,28}$-36$_{10,27}$ & 4.2 & 5.8 & 9.5 & Blend with CH$_3$CHO\\
464.293855 & SO$_2$ & 33$_{5,29}$-33$_{4,30}$ & 4.1 & 10.8 & 5.1 & \\
464.570359 & CH$_2$NH & 7$_{1,6}$-6$_{1,5}$ & 4.4 & 16.8 & 8.5 & \\
464.730426 & U-line & & 12.9 & 6.6 & & \\
464.837513 & CH$_3$OH & 9$_{2}$-9$_{1}$& 34.0 & 11.3 & 7.2 & Blend with C$_2$H$_5$CN 9$_{2}$-9$_{1}$\\
464.928173 & HDO & 1$_{0,1}$-0$_{0,0}$ & 33.5 & 16.2 & 6.6 & \\
465.061222 & C$_2$H$_3$CN & 52$_{2,49}$-49$_{2,48}$ $v_2$ = 1 & 1.9 & 11.1 & 12.3 & \\
465.078910 & U-line & & 3.6 & 4.1 & & \\
465.352350 & $^{34}$SO & N$_{J}$=11$_{12}$-10$_{11}$ & 30.5 & 23.0 & 7.2 & \\
465.512820 & U-line & & 2.4 & 5.4 & & \\
465.543972 & (CH$_3$)$_2$O & 21$_{3,19}$-20$_{2,18}$ & 3.6 & 11.2 & 5.9 & 4 lines blended\\
465.565977 & HCOOCH$_3$ & 36$_{1,35}$-35$_{1,34}$ & 2.4 & 14.7 & 9.0 & \\
465.605514 & C$_2$H$_5$CN & 52$_{12,41/40}$-51$_{12,40/39}$ & 4.0 & 11.2 & 4.4 & \\
465.628191 & SO$_2$ & 29$_{2,28}$-29$_{1,29}$ & 13.0 & 8.4 & 5.1 & \\
465.713762 & C$_2$H$_5$CN & 51$_{5,46}$-50$_{5,45}$ & 7.4 & 13.2 & 5.9 & \\
465.732611 & NH$_2$CHO & 22$_{1,21}$-21$_{1,20}$ & 16.1 & 9.7 & 6.7 & \\
465.758340 & SO$_2$ & 26$_{0,26}$-25$_{1,25}$ & 45.3 & 15.8 & 4.4 & \\
465.882288 & (CH$_3$)$_2$O & 14$_{4,11}$-13$_{3,10}$ & 4.0 & 9.9 & 6.1 & Blend with SO$_2$\\
465.882288 & SO$_2$ & 25$_{10,16}$-26$_{9,17}$+25$_{10,15}$-26$_{9,18}$ & 4.0 & 9.9 & 8.7 & Blended with (CH$_3$)$_2$O\\
465.922779 & SO$_2$ v$_2$ = 1 & 18$_{5,13}$-18$_{4,14}$ & 2.2 & 9.5 & 7.9 & \\
465.956339 & $^{30}$SiO & $J$=11-10 & 7.7 & 15.1 & 8.8 & \\
466.781659 & C$_2$H$_3$CN & 49$_{13,36/37}$-48$_{13,35/36}$ $v_1$ = 1 & 13.8 & 30.7 & 11.9 & Blended with C$_2$H$_5$CN\\
466.781659 & C$_2$H$_5$CN & 52$_{8,44}$-51$_{8,43}$ & 13.8 & 30.7 & 3.8 & Blend with C$_2$H$_3$CN\\
466.892775 & SO$_2$ & 40$_{6,34}$-40$_{5,35}$ & 5.5 & 5.5 & 5.7 & \\
466.999547 & U-line & & 3.1 & 5.6 & & \\
467.048535 & HCOOCH$_3$ & 38$_{23,16/15}$-37$_{23,15/14}$ & 5.6 & 14.7 & 8.9 & \\
\hline \noalign{\smallskip}
\end{tabular}
\end{flushleft}
\label{allines2}
\end{table*}

\begin{table*}
\caption[]{JCMT Orion spectral linesurvey}
\begin{flushleft}
\begin{tabular}{lllllllll}
	\hline Frequency&Species&Transition&Peak&Width&Vel&Notes\\
	\noalign{\smallskip} MHz& & & T$_{mb}$ K& km s$^{-1}$& km s$^{-1}$& \\
	\noalign{\smallskip}
	\hline\noalign{\smallskip}
	\noalign{\smallskip}
	467.212513 & HCOOCH$_3$ & 38$_{22,17}$-37$_{22,16}$ & 4.6 & 5.5 & 8.6 & \\
467.282605 & C$_2$H$_3$CN & 49$_{16,33/34}$-48$_{16,32/33}$ $v_1$ = 1 & 6.9 & 7.6 & 8.2 & \\
467.318607 & HCOOCH$_3$ & 39$_{6,34}$-38$_{6,33}$ & 3.9 & 19.2 & 9.4 & 2 lines blended\\
467.475284 & U-line & & 10.5 & 8.3 & & \\
467.532613 & (CH$_3$)$_2$O & 26$_{0,26}$-25$_{1,25}$ & 6.2 & 5.4 & 7.7 & 4 lines blended\\
467.586402 & HCOOCH$_3$ & 39$_{5,34}$-38$_{5,33}$ & 3.7 & 10.5 & 8.3 & \\
467.627790 & HCOOCH$_3$ & 19$_{9,10}$-18$_{8,10}$ & 2.8 & 10.2 & 8.5 & \\
467.658900 & U-line & & 6.0 & 17.2 & & \\
467.670464 & C$_2$H$_5$OH & 15$_{3,12}$- - 14$_{2,12}$ & 10.4 & 6.4 & 9.4 & \\
467.715623 & (CH$_3$)$_2$O & 26$_{1,26}$-25$_{1,25}$ & 14.4 & 13.0 & 5.6 & 4 lines blended\\
467.765733 & $^{13}$CH$_3$OH & 10$_1$ -- 9$_1$ v$_t$ = 0,1 & 7.1 & 6.4 & & \\
468.048223 & C$_2$H$_5$CN & 55$_{0,55}$-54$_{0,54}$ & 2.9 & 12.8 & 6.5 & \\
468.300479 & CH$_3$OH & 8$_{2}$-8$_{1}$ & 20.3 & 8.4 & 4.7 & Blend with U-line\\
468.345184 & U-line & & 7.6 & 6.0 & & \\
491.943519 & SO$_2$ & 7$_{4,4}$-6$_{3,3}$ & 49.4 & 10.3 & 3.6 & \\
491.978910 & H$_2$CO & 7$_{1,7}$-6$_{1,6}$ & 45.0 & 13.6 & 2.6 &\\
492.160626 & CI & $^3$P$_1$-$^3$P$_1$ & 17.9 & 5.0 & 9.0 & \\
492.281601 & CH$_3$OH & 4$_{1}$-3$_{0}$ & 36.9 & 10.4 & 7.2 & \\
492.470010 & U-line & & 5.2 & 8.5 & & \\
492.544658 & C$_2$H$_5$CN & 55$_{11,45/44}$-54$_{11,44/43}$ & 3.5 & 4.3 & 5.3 & \\
492.695825 & $^{13}$CH$_3$OH & & 9.9 & 4.7 & 8.1 & \\
492.784191 & (CH$_3$)$_2$O & 9$_{6,4}$-8$_{5,4}$  & 20.7 & 8.4 & 8.6 & 8 lines blended\\
493.260821 & C$_2$H$_5$CN & 58$_{1/0,58}$-57$_{1/0,57}$ & 4.4 & 7.4 & 6.2 & \\
493.370372 & U-line & & 9.6 & 11.4 & & \\
493.701318 & CH$_3$OH & 5$_{3}$-4$_{2}$ & 23.7 & 11.4 & 7.6 & \\
493.735068 & CH$_3$OH & 5$_{3}$-4$_{2}$ - & 29.0 & 9.3 & 8.1 & \\
494.460056 & NH$_2$D & 1$_{1,0}$ - 0$_{0,0}$- & 10.5 & 8.6 & 5.8 & \\
494.484560 & CH$_3$OH & 7$_{2}$-7$_{1}$ - & 50.8 & 10.3 & 7.2 & \\
494.555486 & SO$_2$ & 13$_27$-12$_28$ & 2.5 & 4.2 & 7.6 & \\
494.557464 & C$_2$H$_5$CN & 55$_{7,49}$-54$_{7,48}$ & 2.5 & 4.2 & 5.4 & Blend with C$_2$H$_5$OH\\
494.557464 & C$_2$H$_5$OH & 10$_{9,1}$ - 9$_{8,1}$- & 2.5 & 4.2 & 8.7 & Blend with C$_2$H$_5$CN\\
494.575046 & C$_2$H$_5$CN & 32$_{5,27}$-31$_{4,28}$ & 2.6 & 6.4 & 9.5 & \\
494.596465 & (CH$_3$)$_2$O & 28$_{9,19}$-28$_{8,20}$+28$_{9,20}$-28$_{8,21}$ & 5.7 & 5.3 & 7.7 & 8 lines blended\\
494.756403 & C$_2$H$_3$CN & 52$_{17,35/36}$-51$_{17,34/35}$ $v$ = 0 & 14.1 & 17.1 & 5.2 & \\
494.781464 & SO$_2$ & 12$_{3,9}$-11$_{2,10}$ & 45.0 & 26.6 & 7.9 & \\
494.899521 & C$_2$H$_5$OH & 28$_{3,25}$- -27$_{3,24}$- & 1.8 & 10.5 & 3.1 & \\
495.000693 & (CH$_3$)$_2$O & 27$_{9,18}$-27$_{8,19}$+27$_{9,19}$-27$_{8,20}$ & 1.6 & 15.1 & 10.9 & 8 lines blended\\
495.222806 & (CH$_3$)$_2$O & 27$_{1,26}$-26$_{2,25}$ & 7.8 & 14.4 & 7.5 & 4 lines blended\\
495.486885 & HCOOCH$_3$ & 39$_{7,32}$-38$_{7,31}$ & 12.5 & 4.8 & 7.4 & \\
495.821871 & C$_2$H$_5$OH & 26$_{4,22}$-25$_{3,22}$ & 2.1 & 2.8 & 3.5 & \\
495.842389 & SO$_2$ & 13$_{8,6}$-14$_{7,7}$+13$_{8,5}$-14$_{7,8}$ & 9.8 & 20.5 & 8.5 & \\
495.979430 & (CH$_3$)$_2$O & 24$_{9,15}$-24$_{8,16}$+24$_{9,16}$-24$_{8,17}$ & 8.7 & 5.8 & 7.1 & 8 lines blended with CH$_3$CN\\
495.979430 & CH$_3$CN & $J$=27-26 $K$ = 7 & 8.7 & 5.8 & 7.0 & Blend with (CH$_3$)$_2$O\\
496.107704 & CH$_3$CN & $J$=27-26 $K$ = 6 & 11.5 & 11.6 & 4.6 & \\
496.207453 & CH$_3$CN & $J$=27-26 $K$ = 5 & 8.8 & 13.3 & 6.6 & \\
496.295793 & CH$_3$CN & $J$=27-26 $K$ = 4 & 13.3 & 11.1 & 4.8 & \\
496.361787 & CH$_3$CN & $J$=27-26 $K$ = 3 & 19.2 & 11.6 & 5.1 & \\
496.409463 & CH$_3$CN & $J$=27-26 $K$ = 2 & 20.7 & 12.5 & 5.0 & \\
496.440950 & (CH$_3$)$_2$O & 24$_{9,15}$-24$_{8,16}$+24$_{9,16}$-24$_{8,17}$ & 26.6 & 17.0 & 10.9 & 8 lines blended with CH$_3$CN\\
495.175350 & CH$_3$OH & 7$_{0}$-6$_{-1}$ & 39.7 & 10.2 & 7.7 & \\
496.440950 & CH$_3$CN & $J$=27-26 $K$ = 1 & 26.6 & 17.0 & 3.2 & 2 lines blended, Blend with (CH$_3$)$_2$O\\
496.518825 & H$_2$$^{13}$CO & 7$_{2,6}$-6$_{2,5}$ & 3.1 & 7.2 & 6.3 & \\
496.633861 & (CH$_3$)$_2$O & 21$_{9,12}$-21$_{8,13}$+21$_{9,13}$-21$_{8,14}$ & 7.0 & 5.2 & 8.4 & 8 lines blended\\
496.764492 & C$_2$H$_3$CN & 52$_{5,47}$-51$_{5,46}$ $v_2$ = 1 & 2.2 & 14.4 & 14.1 & \\
496.796267 & (CH$_3$)$_2$O & 20$_{9,11}$-20$_{8,12}$+20$_{9,12}$-20$_{8,13}$ & 5.8 & 3.8 & 8.4 & 8 lines blended\\
496.928524 & CH$_3$OH & 14$_{0}$-13$_{1}$ & 16.7 & 9.2 & 6.5 & \\
497.051216 & (CH$_3$)$_2$O & 18$_{9,9}$-18$_{8,10}$+18$_{9,10}$-18$_{8,11}$ & 7.2 & 14.8 & 8.6 & 8 lines blended\\
\hline \noalign{\smallskip}
\end{tabular}
\end{flushleft}
\label{allines3}
\end{table*}

\begin{table*}
\caption[]{JCMT Orion spectral linesurvey}
\begin{flushleft}
\begin{tabular}{lllllllll}
	\hline Frequency&Species&Transition&Peak&Width&Vel&Notes\\
	\noalign{\smallskip} MHz& & & T$_{mb}$ K& km s$^{-1}$& km s$^{-1}$& \\
	\noalign{\smallskip}
	\hline\noalign{\smallskip}
	\noalign{\smallskip}
497.085708 & C$_2$H$_5$OH & 9$_{7,3/2}$-8$_{6,2/3}$ & 3.1 & 10.4 & 10.3 & \\
497.119025 & H$_2$$^{13}$CO & 7$_{4,4/3}$-6$_{4,3/2}$ & 4.8 & 6.9 & 6.6 & \\
497.149132 & (CH$_3$)$_2$O & 17$_{9,8}$-17$_{8,9}$+17$_{9,9}$-17$_{8,10}$ & 6.4 & 7.3 & 8.5 & 8 lines blended\\
497.226494 & (CH$_3$)$_2$O & 16$_{9,7}$-16$_{8,8}$+16$_{9,8}$-16$_{8,9}$ & 5.5 & 4.8 & 10.4 & 8 lines blended\\
497.295820 & (CH$_3$)$_2$O & 15$_{9,6}$-15$_{8,7}$+15$_{9,7}$-15$_{8,8}$ & 5.2 & 3.2 & 8.1 & 4 lines blended\\
497.364091 & U-line & & 3.7 & 7.0 & & \\
497.388286 & (CH$_3$)$_2$O & 13$_{9,4}$-13$_{8,5}$+13$_{9,5}$-13$_{8,6}$ & 8.5 & 10.8 & 8.4 & 8 lines blended\\
497.404849 & U-line & & 4.4 & 1.5 & & \\
497.417697 & (CH$_3$)$_2$O & 12$_{9,3}$-12$_{8,4}$+12$_{9,4}$-12$_{8,5}$ & 9.4 & 8.4 & 9.4 & 8 lines blended\\
497.442262 & (CH$_3$)$_2$O & 11$_{9,2}$-11$_{8,3}$+11$_{9,3}$-11$_{8,4}$ & 4.2 & 9.4 & 8.3 & 8 lines blended with HCOOCH$_3$\\
497.442262 & HCOOCH$_3$ & 40$_{13,28}$-39$_{13,27}$ +40$_{15,26}$-39$_{13,27}$ & 4.2 & 9.4 & 9.0 & 2 lines blended with (CH$_3$)$_2$O\\
497.516476 & C$_2$H$_3$CN & 60$_{1,59}$-60$_{1,60}$ $v$ = 0 & 4.8 & 6.1 & 6.8 & \\
497.556527 & U-line & & 1.5 & 9.4 & & \\
497.606988 & (CH$_3$)$_2$O & 24$_{3,22}$-23$_{2,21}$ & 4.3 & 2.7 & 8.5 & 4 lines blended\\
497.661575 & C$_2$H$_5$CN & 56$_{3,53}$-55$_{3,52}$ & 4.9 & 6.0 & 3.5 & Blend with CH$_3$CN $v$ = 8\\
497.661575 & CH$_3$CN $v$$_8$=1 & $J$=27-26 K=4 $_{-l}$ & 4.9 & 6.0 & 4.9 & Blend with C$_2$H$_5$CN\\
497.734975 & CH$_3$CN $v$$_8$=1 & $J$=27-26 K=3 $_{-l}$ & 3.5 & 4.5 & 8.8 & \\
497.828685 & CH$_3$OH & 8$_{2}$-8$_{1}$ - & 14.0 & 9.4 & 8.7 & \\
497.905738 & CH$_3$CN $v$$_8$=1 & $J$=27-26 K=4 $_{+l}$ & 3.6 & 6.4 & 5.9 & \\
498.288253 & C$_2$H$_5$CN & 57$_{2,55}$-56$_{2,54}$ & 2.2 & 6.6 & 5.1 & \\
498.327509 & OCS & $J$=41-40 & 12.6 & 10.2 & 6.6 & \\
498.412959 & CH$_3$CN $v$$_8$=1 & $J$=27-26 K=1 $_{+l}$ & 6.4 & 6.8 & 6.9 & \\
498.431868 & $^{34}$SO$_2$ & 8$_{4,4}$-7$_{3,5}$ & 5.9 & 12.2 & 10.0 & \\
498.596075 & HCOOCH$_3$ & 41$_{7,35}$-40$_{7,34}$ & 2.2 & 8.1 & 9.2 & \\
498.677131 & $^{13}$CH$_3$OH & 8$_2$-8$_1$& 8.5 & 7.3 & 4.8 & \\
498.785157 & C$_2$H$_5$CN & 55$_{4,51}$-54$_{4,50}$ & 2.0 & 9.8 & 4.5 & \\
498.849325 & (CH$_3$)$_2$O & 16$_{4,13}$-15$_{3,12}$ & 7.7 & 7.0 & 8.5 & 4 lines blended with HCOOCH$_3$\\
498.849325 & HCOOCH$_3$ & 42$_{5,37}$-41$_{5,36}$ & 7.7 & 7.0 & 6.8 &Blend with (CH$_3$)$_2$O\\
498.984855 & C$_2$H$_5$CN & 55$_{6,49}$-54$_{6,48}$ & 5.4 & 23.7 & 12.0 & \\
499.198900 & HCOOCH$_3$ & 43$_{5,39}$-42$_{5,38}$ & 3.0 & 6.5 & 8.1 & \\
499.283980 & C$_2$H$_3$CN & 20$_{4,17}$-19$_{3,16}$ $v_1$ = 1 & 3.4 & 13.9 & 7.3 &Blend with C$_2$H$_5$CN\\
499.283980 & C$_2$H$_5$CN & 17$_{8,10/9}$-16$_{7,9/10}$ & 3.4 & 13.9 & 3.4 &Blend with C$_2$H$_3$CN\\
499.367702 & (CH$_3$)$_2$O & 27$_{2,26}$-26$_{1,25}$ & 3.9 & 6.1 & 7.1 & 4 lines blended\\
499.892902 & C$_2$H$_5$CN & 58$_{1/2,57}$-57$_{1/2,56}$ & 9.1 & 8.0 & 5.4 & \\
499.917085 & NH$_2$CN & 25$_{2,23}$-24$_{2,22}$ & 8.1 & 8.1 & 8.2 & $^{13}$C$^{34}$S J=11--10 ?\\
499.944145 & U-line & & 10.2 & 9.6 & & \\
500.038140 & HC$_3$N & $J$=55-54 & 9.0 & 7.9 & 6.8 & \\
500.052075 & NH$_2$CN & 25$_{3,22}$-24$_{3,21}$ & 8.0 & 11.8 & 4.9 & \\
500.075688 & SO$_2$ v$_2$=1 & 7$_{4,4}$-6$_{3,3}$ & 3.8 & 11.5 & 8.6 & \\
500.364813 & C$_2$H$_3$CN & 41$_{6,35}$-41$_{5,36}$ $v_2$ = 1 & 4.9 & 5.7 & 6.8 & \\
500.432652 & SO$_2$ & 18$_{9,9}$-19$_{8,12}$+18$_{9,10}$-19$_{8,11}$ & 20.1 & 6.0 & 8.3 & \\
500.457324 & C$_2$H$_5$CN & 12$_{9,3/4}$-11$_{8,4/3}$ & 22.3 & 8.2 & 2.9 & \\
500.475691 & CH$_2$NH & 2$_{2,1}$-2$_{1,2}$ & 18.0 & 7.5 & 6.3 & \\
500.637890 & C$_2$H$_5$CN & 35$_{5,31}$-34$_{4,30}$ & 5.9 & 5.1 & 5.5 & \\
500.661562 & SO$_2$ & 31$_{2,30}$-31$_{1,31}$ & 9.0 & 13.3 & 5.3 & \\
501.087663 & C$_2$H$_3$CN & 55$_{1,55}$-54$_{1,54}$ $v_1$ = 1 & 11.7 & 12.8 & 10.4 & \\
501.116812 & SO$_2$ & 28$_{0,28}$-27$_{1,27}$ & 21.5 & 19.1 & 3.5 & \\
501.592173 & CH$_3$OH & 9$_{2}$-9$_{1}$ - & 32.9 & 11.7 & 7.1 & \\
501.656920 & C$_2$H$_5$CN & 59$_{1/0,59}$-58$_{1/0,58}$ & 5.6 & 8.1 & 8.2 & \\
501.679997 & CH$_2$NH & 8$_{0,8}$-7$_{0,7}$ & 4.4 & 11.5 & 9.0 & \\
501.891221 & (CH$_3$)$_2$O & 30$_{3,27}$-29$_{4,26}$ & 2.8 & 12.0 & 9.1 & 4 lines blended\\
502.181627 & HCOOCH$_3$ & 15$_{12,3/4}$-14$_{11,4/3}$ & 4.6 & 7.0 & 8.2 & \\
502.303684 & N$_2$O & $J$=20-19 & 3.4 & 17.4 & 4.7 & \\
503.018652 & (CH$_3$)$_2$O & 28$_{0,28}$-27$_{1,27}$ & 10.4 & 10.7 & 7.4 & 4 lines blended\\
503.107069 & (CH$_3$)$_2$O & 28$_{1,28}$-27$_{0,27}$ & 5.1 & 5.2 & 9.2 & 4 lines blended\\
503.210868 & HCOOCH$_3$ & 40$_{8,32}$-39$_{8,31}$ & 2.7 & 7.6 & 8.5 & \\
503.518307 & HCOOCH$_3$ & 22$_{9,13}$-21$_{8,14}$ & 6.2 & 6.1 & 7.6 & \\
\hline \noalign{\smallskip}
\end{tabular}
\end{flushleft}
\label{allines4}
\end{table*}

\begin{table*}
\caption[]{JCMT Orion spectral linesurvey}
\begin{flushleft}
\begin{tabular}{lllllllll}
	\hline Frequency&Species&Transition&Peak&Width&Vel&Notes\\
	\noalign{\smallskip} MHz& & & T$_{mb}$ K& km s$^{-1}$& km s$^{-1}$& \\
	\noalign{\smallskip}
	\hline\noalign{\smallskip}
	\noalign{\smallskip}
503.574370 & C$_2$H$_3$CN & 53$_{14,39/40}$-52$_{14,38/39}$ $v$ = 0 & 9.5 & 5.6 & 7.9 & Blend with CH$_3$OH\\
503.574370 & CH$_3$OH & 7$_{6}$-8$_{5}$ & 9.5 & 5.6 & 7.5 & Blend with C$_2$H$_3$CN\\
503.707705 & HCOOCH$_3$ & 41$_{26,16}$-40$_{26,15}$ & 3.3 & 6.8 & 8.4 & \\
503.836578 & C$_2$H$_5$CN & 28$_{6,23}$-27$_{5,22}$ & 3.4 & 2.1 & 6.8 & \\
503.851690 & SO & N$_{J}$=15$_{14}$-14$_{14}$ & 8.3 & 10.9 & 2.1 & Blend with C$_2$H$_3$CN\\
503.921239 & C$_2$H$_3$CN & 29$_{6,24}$-29$_{5,25}$ $v_2$ = 1 & 2.0 & 7.9 & 12.5 & \\
504.297599 & CH$_3$OH & 7$_{1}$-6$_{0}$ & 34.3 & 8.9 & 6.6 & \\
504.413582 & C$_2$H$_3$CN & 23$_{6,17/18}$-23$_{5,18/19}$ $v_2$ = 1 & 2.4 & 16.2 & 7.6 & \\
504.510971 & SO$_2$ & 23$_{10,14}$-24$_{9,15}$+23$_{10,13}$-24$_{9,16}$ & 2.4 & 25.3 & 9.5 & \\
504.680057 & SO & N$_{J}$=4$_{3}$-1$_{2}$ & 9.3 & 26.7 & 6.8 & \\
504.728059 & $^{34}$SO & N$_{J}$=12$_{11}$-11$_{10}$ & 20.5 & 22.4 & 6.6 & Blend with C$_2$H$_5$CN\\
504.728059 & C$_2$H$_5$CN & 57$_{4,54}$-56$_{4,53}$ & 20.5 & 22.4 & -0.6 & Blend with $^{34}$SO\\
505.431404 & U-line & & 2.7 & 7.7 & & \\
505.499853 & C$_2$H$_5$OH & 29$_{6,23}$+28$_{6,22}$& 3.5 & 7.9 & 9.3 & \\
505.765740 & CH$_3$OH & 10$_{2}$-10$_{1}$ & 31.4 & 10.3 & 6.8 & \\
505.838107 & H$_2$CO & 7$_{0,7}$-6$_{0,6}$ & 47.9 & 17.5 & 6.4 & \\
506.157737 & CH$_3$OH & 11$_{1}$-10$_{2}$ & 26.0 & 8.3 & 6.3 & \\
506.240528 & HCOOCH$_3$ & 41$_{17,25/24}$-40$_{17,24/23}$+41$_{17,25/24}$-40$_{17,23/24}$ & 13.7 & 25.5 & 6.5 & 4 lines\\
506.253692 & $^{34}$SO & N$_{J}$=12$_{12}$-11$_{11}$ & 11.0 & 16.9 & 7.6 & \\
506.773898 & U-line & & 6.3 & 10.0 & & \\
506.831598 & DCN & $J$=7-6 & 8.7 & 11.8 & 6.6 & \\
507.041856 & U-line & & 4.5 & 5.9 & & \\
507.200247 & HNCO & 23$_{1,22}$-22$_{1,21}$ & 9.6 & 10.4 & 6.1 & \\
507.301540 & $^{34}$SO & N$_{J}$=12$_{13}$-11$_{12}$ & 24.9 & 27.8 & 9.0 & \\
507.337692 & SO$_2$ & 36$_{6,30}$-35$_{5,31}$ & 4.3 & 6.9 & 4.8 & \\
507.383231 & C$_2$H$_3$CN & 27$_{3,24}$-26$_{2,25}$ $v_1$ = 1 & 8.7 & 6.6 & 6.1 & \\
\hline \noalign{\smallskip}
\end{tabular}
\end{flushleft}
\label{allines5}
\end{table*}

\subsection{CO}

In this survey the CO $J$ = 4--3 transition is the most intense single line, with a full width at half maximum of 40 \kmsec and full width at zero intensity of at least 120 \kmsec. The peak main beam brightness temperature \Tmb = 182 K was similar to values that we have previously measured using the JCMT (see for example White \& Sandell 1995). The line profile clearly shows a small dip close to its peak,
which is probably a self-absorption dip. This has also been reported by Schilke \al (2001) in the $J$ = 6--5 transition, and was seen in all of our earlier unpublished JCMT spectra in this transition using both the present SIS receiver, as well as the original JCMT 460 - 490 GHz Indium Antimonide receiver (White $\&$ Padman 1991, Padman $\&$ White 1992). We have made careful checks on the many occasions
that we have observed this line with the JCMT. confirming that this dip is not a result of the subtraction
of emission located at the off position. Unpublished maps (in preparation) of the spatial
distribution of this absorption feature show it to be spatially localised on the hot core, and that larger scale CO emission is spatially extended around the core - meaning that the \Tmb may overestimate the true kinetic temperature.

\subsection{Sulphur Monoxide (SO) and Sulphur Dioxide (SO$_{2}$)}

The high abundances of sulphur based molecules in the interstellar medium are believed to be in part due to the presence of shocks, that favour the endothermic reactions required to form these molecules. The rotational temperature diagram for SO$_2$ (see Fig \ref{rotemp}) is consistent with previous estimates of the temperature and \Ncol. The temperature is 136 K, typical of the cooler conditions found in the plateau. The value of \Ncol is estimated to be 9.7 10$^{16}$ \cmtwo in the optically thin limit, increasing to 1.2 10$^{17}$ \cmtwo after making a correction for optical depth. Since our initial assumptions might be that this molecule should almost certainly be moderately optically thick, we ran a model, based on the Sutton \al (1995) temperature and column density estimate, calculating the expected optical depths for the 35 SO$_2$ lines observed in this survey. One transition has a strong opacity $\tau$ = 1.46. The optical depth of all other lines are less than 0.9 and for half of $all$ the detected lines $\tau$ $\leq$ 0.2. Therefore the results of the optically thick and optically thin estimates of column density are understandably close to each other. This result is in reasonable agreement with studies using similar beam sizes, by Schilke \al (2001) (\Ncol = 9.7 10$^{16}$ \cmtwo, $T$ $\sim$ 187 K for a 12 - 14\arcsec beam) and Sutton \al (1995) (\Ncol = 9.4 10$^{16}$ \cmtwo, $T$ $\sim$ 99 K for a 12 - 14\arcsec beam).

We further estimate that the isotopic ratio [SO$_2$]/[$^{34}$SO$_2$] = 14.1, which is in agreement with previous estimates by Blake \al (1987) (14 - 16) and Schloerb \al 1983 (11).
\label{so2section}

\subsection{Methyl Cyanide (CH$_{3}$CN), Ethyl Cyanide (propionitrile) (C$_{2}$H$_{5}$CN), Vinyl Cyanide (acrylonitrile) C$_{2}$H$_{3}$CN)}
The $J$ = 25--24 and $J$ = 27--26 lines of Methyl Cyanide were observed at 459 and 495 GHz respectively. The average LSR velocity of the lines were $\sim$ 5.6 \kmsec and the line widths $\sim$11 \kmsec. It is therefore likely that this species traces the hot core. The $J$ = 25--24 $k$ = 7 line (459.276 GHz) could not be assigned due to a strong U-line (459.267 GHz). Lines from the $J$ = 27--26 $k$ = 2 (-1) and $k$ = 3 (+1) transitions were observed at 497.790 and 497.971 GHz respectively, although they were not used in the analysis due to their low intensities. The \Ncol and \Trot values for this molecule were $3.5\times 10^{15}$ cm$^{-2}$ and $\sim$ 230 K - similar to those estimated by Wilner \al 1994, Sutton \al 1995, Lee, Cho $\&$ Lee (2001). Two lines of the isotopomeric variant CH$_{3}$$^{13}$CN were also detected. No lines from the $v_8$ = 1 family were identified.

For this molecule, a) \Trot was fixed at 227 K, obtained from analysis of the CH$_{\rm 3}$CN line, and b) \Trot and \Ncol were calculated from the two observed lines. In this latter case, the value of \Trot is reduced to about one third that of CH$_{\rm 3}$CN. We consider that case a) provides a better solution, because of the difficulty of estimating \Trot from two lines whose upper state energy levels are very similar to each other. It did not prove possible to identify the $J$ = 25--24 $k$ = 0, 1, 3 lines due to blending with CH$_{3}$CN. In view of the problems that have previously been encountered in the interpretation of CH$_3$CN lines using an optically thin assumption (Schilke \al 1999, Comito \al 2003), we ran a model to calculate the opacity of the various lines based on the above excitation conditions estimated by Wilner \al 1994 (obtained with a similar beam size to our study). The largest value of $\tau$ amongst the present lines is estimated to be 0.2, and all other values were $\leq$ 0.1. Making a correction for the optical depth only increases the value of column density by a few percent - from \Ncol = 3.5 10$^{15}$ \cmtwo to 3.6 10$^{16}$ \cmtwo. We therefore conclude that with few exceptions, the optically thin assumption provides a valid and useful estimate of column density for the transitions observed in this survey. 

Thirteen lines of Vinyl Cyanide and its isotopomeric variant C$_{\rm 2}$H$_{\rm 3}$CN were detected at LSR velocities consistent with the molecules being concentrated in the compact ridge. Line frequencies have been reported by Demaison $\al$ (1994). 5 lines were tentatively associated with the $v$$_{15}$ (out-of-plane bend) = 1 vibrational excited state. However, since the frequencies of the $v$$_{15}$ lines have a large error, they were not included in the fitting. We also searched at the expected frequency of the $v$$_{11}$ = 2 state, but could not find convincing match. Grain surface reactions are thought to be the main process by which complex nitrogen bearing species are formed, due the high level of hydrogenation. It is not likely that a molecule such as CCCN would pick up the required number of hydrogen atoms to become ethyl cyanide without the intermediate step of adsorption on to a grain, allowing the hydrogenation process to occur. The high temperatures found in the hot core are sufficient to evaporate the molecules from the surface of any grains that drift into the region.

Twenty seven lines of Ethyl Cyanide were detected, based on line frequencies taken from Pearson \al (1997) and Pearson (2000). It is expected that many observed lines of this molecule are
blended with U-lines, and it was found that the simultaneous estimation of the column density and rotational temperature were difficult, since many of the lines are located very close together on the rotational diagram. To avoid this problem. we have calculated the column densities of C$_{\rm 2}$H$_{\rm 5}$CN for two fixed temperatures, deriving values of \Ncol = 2.4 10$^{16}$ \cmtwo (assuming a temperature of 150K based on the work of Sutton \al (1985), or \Ncol = 8.310$^{15}$ \cmtwo (assuming a temperature of 239K based on the work of Schilke \al (2001). As a result, the best estimates of the column density of the C$_{\rm 2}$H$_{\rm 5}$CN from the present data are in the range \Ncol = 2.4--0.83 10$^{16}$ \cmtwo. This value is close to that estimated by Schilke \al (2001) (\Ncol = 3.1 10$^{16}$ \cmtwo) with a beam size of 10-12\arcsec - which is very similar to that of the present survey. By comparison,  Sutton \al (1985) reported \Ncol = 2 10$^{15}$ \cmtwo with a 30\arcsec beam. Assuming that the emitting region is a core with 10\arcsec diameter, \Ncol = 1.5 10$^{16}$ \cmtwo, which is in better agreement with the estimate in the present survey value. Similar arguments applied to the Methyl Cyanide line, to resolve the differences between the column densities from this survey, and the work of Blake \al (1985) and Schilke \al (2001), again suggest that the size of the emitting region is $\sim$ 10\arcsec.

\label{ch3cnsection}

\subsection{Cyanamide (NH$_{\rm 2}$CN)}
Three lines of NH$_{\rm 2}$CN were detected. However, it did not prove possible to obtain a reliable \Trot from the observed data. Consequently we assumed values of \Trot of 100 and 200K, to make a first estimate of the column densities.

\subsection{Carbonyl sulphide (OCS)}
Two lines of OCS were detected. The fit for \Trot and \Ncol are not very accurate, since the upper state energy levels of the two lines are very similar.

\subsection{Methylenimeme (CH$_{2}$NH)}
Methylenimine is a prolate asymmetric rotor with components of the electric dipole moment along both the $a$ and $b$ molecular axes, with magnitudes 1.325 and 1.53 D, respectively. The nitrogen nucleus produces electric quadrupole hyperfine structure in low-lying transitions. CH$_{\rm 2}$NH is a likely product following the UV irradiation of icy interstellar grain mantles and may be a precursor of other complex organics which may be present in cometary ices (Bernstein \al 1995), and may be a precursor to glycinenitrile and glycine (e.g., Dickerson 1978). Three lines of CH$_{\rm 2}$NH were used in the fitting. It was considered likely that the 2$_{2,1}$ -- 2$_{1,2}$ line was probably blended with a U-line, since the intensity was strong compared with the other two lines.

Using only two lines, it did not prove possible to get a reasonable value for \Trot. Hence, in fitting the value of \Trot was fixed at 150 K, from observations of HNCO, which has a similar dipole moment (CH$_{\rm 2}$NH $\mu$(a) = 1.352 D, $\mu$(b) = 1.530 D, HNCO $\mu$(a) = 1.602 D, $\mu$(b) = 1.35 D - from Kirchoff, Johnson $\&$ Lovas 1973, Hocking \al 1975). The derived column density, \Ncol = 2.4 10$^{15}$ $\pm$1.8 10$^{14}$ \cmtwo, is in reasonable agreement with the column density reported in the detection paper by Dickens \al (1977) of \Ncol = 6.2 $\pm$ 1,8 10$^{14}$ \cmtwo, based on observations of 5 transitions (three of which were blended with other lines) between 172 and 256 GHz obtained with a larger beam of 23 - 34\arcsec.

\subsection{Methanol (CH$_{\rm 3}$OH) and Ethanol (C$_{\rm 2}$H$_{\rm 5}$OH)}
Methanol, CH$_{\rm 3}$OH, is one of the most widely observed molecules in star forming regions. Line frequencies are reported by Xu $\&$ Lovas (1997). Twenty four CH$_{\rm 3}$OH lines were detected with an average velocity of 7.8 \kmsec and line width of 9.9 \kmsec - suggesting that it is likely to be excited in the compact ridge. The high observed abundances  of this molecule imply a high abundance of the precursor ion CH$_{\rm 3}^{+}$ which would react with water in a ion-molecule process to form methanol. Two lines of $^{\rm 13}$CH$_{\rm 3}$OH (which were apparently detected at 492.695 and 498.677 GHz) would indicate that the main methanol line may have an opacity of at least 0.7. Schilke \al (2001) have pointed out the difficulty of applying simple rotational analysis techniques to methanol due to opacity problems. The column densities reported in Table 1 have been should therefore be treated as lower limits. If the 498.677 GHz line is assigned to $^{\rm 13}$CH$_{\rm 3}$OH-A(8$_2$--8$_1$), the column density of this molecule would be 8.7 10$^{15}$ cm$^{-2}$ assuming the rotational temperature is the same as that of $^{\rm 12}$CH$_{\rm 3}$OH (600 K - but note that the error on this is 50\%). This would indicate that the [$^{12}$C]/[$^{13}$C] ratio $\sim$ 10 - although this would also decrease if the temperature were substantially less than 600K). Neither the $^{\rm 13}$CH$_{\rm 3}$OH--E(23$_-2$-23$_-1$) line at 503.216 GHz, nor any lines of torsionally excited methanol were identified.

Eight lines of C$_{\rm 2}$H$_{\rm 5}$OH were detected, at a velocity of $\sim$ 8 \kmsec, although only 6 were used in the rotational temperature fitting. Line frequencies are given by Pearson $\al$ (1996). The lines at 494.899 and 495.821 had line widths that were too broad and narrow respectively, and it is likely that these are blended, or due to other species. For this molecule, we made two fits; a) simultaneously fitting both \Ncol and \Trot, and b) \Trot was fixed at 70 K (following Ohishi \al 1995), since simultaneous fitting of both \Trot and \Ncol led to large uncertainties, and a large \Trot.

\subsection{Methyl formate (HCOOCH$_{\rm 3}$)}
The line widths and velocities of the HCOOCH$_{\rm 3}$ lines suggest an origin in the compact ridge. Line frequencies are given by Osterling $\al$ (1999). Lines are seen from both the E- and the A-symmetry states. The high abundance follows directly from the high abundance of methanol whose precursor ion CH$_{\rm 3}$OH$_{\rm 2}^{+}$ reacts with H$_{\rm 2}$CO. The column density we derive is similar to that estimated by Schilke \al (1997), who have discussed difficulties in assigning an unique rotation temperature for this line.

\subsection{Dimethyl ether ((CH$_{\rm 3}$)$_{\rm 2}$O)}
Dimethyl Ether, (CH$_{\rm 3}$)$_{\rm 2}$O, is one of the few interstellar molecules whose emission lines are affected by the presence of two internal rotors. Line frequencies are reported by Groner $\al$ (1998). Twenty seven lines were detected in this survey, which were best described by a fit of \Ncol = 1.4 10$^{16}$ \cmtwo and \Trot = 157 K for the column density and rotational temperature respectively. These values are close to those found by Sutton et al (1985) toward the compact ridge. The line widths and velocities are consistent with an origin in the compact ridge. The fact  that this molecule appears to be far more abundant (from the large number of  strong lines) than ethanol leads to the conclusion that the Compact ridge,  where these molecules are formed, is rich in hydrogen. The molecule is formed  by a similar method to methyl formate except that  CH$_{\rm 3}$OH$_{\rm 2}^{+}$ reacts with methanol to form it.

\subsection{Formaldehyde (H$_{\rm 2}$CO)}
H$_{\rm 2}$CO is a highly prolate asymmetric top molecule. The rotational temperature of H$_{\rm 2}$CO was fixed at 166 K following Blake \al (1987). Two lines were associated with this molecule, however an additional line at 491.979 GHz was not included in the fitting, since its velocity appeared to be low (2.6 \kmsec). The derived column density is very similar to that estimated by Blake \al (1987).

\subsection{Deuterated water (HDO)}

A single transition of HDO was detected at 464.925 GHz. It has a clear hot 
core line shape with a line width of 16.2 \kmsec and a velocity of 6.6 \kmsec. We have attempted to model this line along with the two other lines reported and analysed by Schulz \al (1991) and Sutton \al (\cite{sutton:etal}). Pardo \al (2001b) have recently reported detections of the $J_{\rm Ka,Kb}$ = 2$_{\rm 1,2}$ -- 1$_{\rm 1,1}$ and 1$_{\rm 1,1}$ -- 0$_{\rm 0,0}$ lines in the 850 - 900 GHz region, which appear trace the plateau gas, rather than the hot core material which is contributes rather more to the other HDO lines observed to date. We therefore estimate values of 3.2$\times$ 10$^{16}$ cm$^{-2}$ and 164 K for \Ncol and \Trot respectively, in broad agreement with values reported by Plambeck \& Wright (1987) and Jacq \al (1990). This is as expected from the high abundances observed for molecules such as methanol and methyl formate which require water to be present for their formation. It is thought that both water and deuterated water are formed elsewhere on grains and evaporated from the surfaces in the higher temperature conditions found in the hot core (Beckman \al 1981, Pardo \al 2001b).

\subsection{Atomic Carbon (CI)}
A lower limit on the column density of the \3P1 transition was determined with the treatment of the optically thin line emission given by White $\&$ Sandell (1995). The column density, \Ncol = 1.2 10$^{18}$ \cmtwo is consistent with that of White $\&$ Sandell (1995). Keene \al (1998) detected the $^{13}$CI line toward a position 4\arcmin south of our pointing position, showing that the main isotopomeric line was optically thin. To date, this remains the best indicator that it is useful to use the optically thin column density estimate.

\subsection{Other Molecules}
Other detections that are of interest include HCCCN and HNCO (Isocyanic Acid), both with line shapes are typical of the hot core, and with estimated abundance similar to those reported by Schilke \al (1997).

\subsection{Unidentified Lines}

At total of 33 lines were detected that could not be associated with known spectral lines. We searched carefully in the Cologne Database for Molecular Spectroscopy (M$\ddot{u}$ller \al 2001) and the JPL Molecular Spectroscopy Database, as well as other tables of isotopomeric variants of lines that were present in the survey (e.g. CH$_{\rm 3}$OH and $^{\rm 13}$CH$_{\rm 3}$OH from the laboratory measurements of Anderson, Herbst $\&$ de Lucia 1987, 1990, 1992). Although there were inevitably a number of lines that lay close to some of the U-line frequencies, we also used secondary criteria (line strength predictions from the Cologne database, upper energy state levels, and the presence or absence of other lines from similar levels) to make judgments as to whether lines could be associated with particular species. Six lines that were originally designated as U-lines were associated in this way, however a substantial number of intense lines remain. Assignment of lines to these frequencies is beyond the scope of this paper, and will require a sophisticated modeling effort, combined with U-lines from other published surveys. We note that the number of U-lines inferred from the present survey (33 - or 13\% of the total) is similar reported from some other surveys (325 - 360 GHz $\sim$ 8\% Sutton \al 1995), 607 - 725 GHz $\sim$ 14\% (Comito \al 2003).

\subsection{Notes on \Trot diagrams}
\begin{figure}
\resizebox{\hsize}{!}{\includegraphics{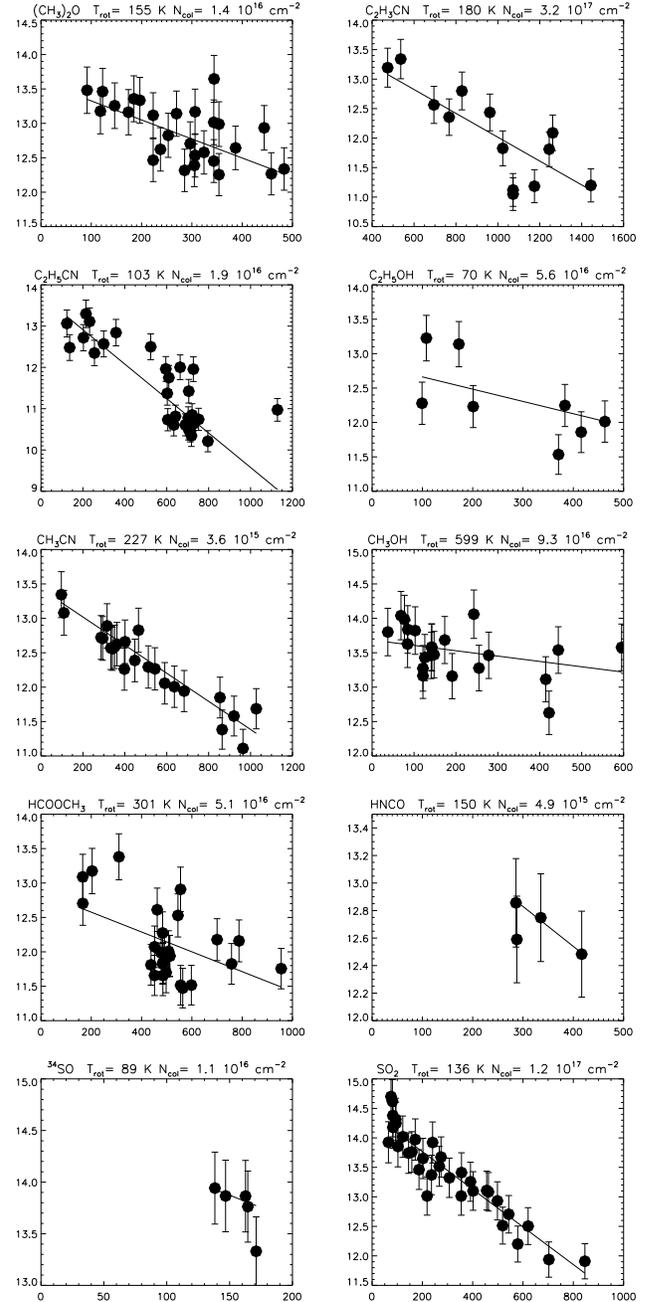}}
\caption{<Rotational temperature diagrams. The fitted lines were derived from a noise-weighted least squares fitting procedure.}
\label{rotemp}
\end{figure}

Plots used to fit the estimates of \Trot or \Ncol are shown in Fig. \ref{rotemp}.

The values for the intrinsic line strengths, level degeneracies and partition functions are derived from the literature, including the spectral line catalogues cited previously (see also Eq 1 and 2).

\subsection{Line-to-continuum ratio}
The data may also be used to estimate the ratio of line-to-continuum emission from the source. This is an important value to characterise for a range of molecular clouds - since a considerable fraction of the fluxes measured with submm continuum cameras, such as the JCMT's SCUBA, and the CSO's SHARC, may be due to the integrated line emission, rather than thermal dust. Early attempts to measure, or characterise this ratio toward Orion have included the work of Groesbeck (1995), Greaves $\&$ White (\cite {greaves:white}), White $\&$ Sandell (\cite{white:sandell}), Schilke \al 1997, 2001). The integrated line emission of the spectrum shown in Fig \ref{composite} is $\sim$ 1.1 $\times$ 10$^{5}$ K MHz. The $average$ rms noise level  over the whole survey range was 1.5 K per 2 MHz channel, and therefore line emission would not be detected under this level. If lines of this strength were uniformly spread across the 30 GHz spectral range, then this would contribute a further $\sim$ 4.5 $\times$ 10$^{4}$ K MHz to the integrated line emission. The total estimated integrated emission corresponds to an integrated main beam temperature of 3.7 K MHz$^{-1}$. Greaves $\&$ White (\cite {greaves:white}) reported an integrated emission of $\sim$ 1.0 $\times$ 10$^{4}$ K MHz over a total range of 16 GHz, which corresponded to an average temperature of 0.6 K MHz$^{-1}$ - or about one sixth of that estimated here. This discrepancy can partially be explained by the fact that the hot core is smaller in angular extent that the beam size used in both this and the survey of Greaves $\&$ White (\cite {greaves:white}). The coupling efficiency $\eta_{c}$ to a Gaussian source of radius  $\sigma_{s}$ is given by Eq. \ref{beamdil}:

\begin{equation}
\eta_{c}=\frac{\sigma_{s}^{2}}{\left(\sigma_{s}^{2}+\sigma^{2}\right)}
\label{beamdil}
\end{equation}

where  $\sigma$ is the Gaussian radius of the telescope beam. Taking $\sigma_{s}$ = 4\arcsec and $\sigma = 4.2$\arcsec (for a 10\arcsec FWHM beam), $\eta_{c} = 0.48$, and $\eta_{c} = 0.25$ for the survey of Greaves $\&$ White (\cite{greaves:white}) would lead to an expectation that the current value should be double that previously estimated for this reason. The lower beam dilution in this survey could explain the rest of the discrepancy. No measurements of the continuum emission exist for the exact frequency that this survey was conducted at, but by using the measurements made by White $\&$ Sandell (\cite{white:sandell}) at 790 \mum it is possible to extrapolate and make an estimate of the emission at 600 \mum (500 GHz). The spectral index was estimated by White $\&$ Sandell (\cite{white:sandell}) to be 1.5 and the emission was 134.9 Jy in a 10\arcsec beam with an aperture efficiency of 0.3.  The antenna temperature ($T_{A}$) is given by Eq. \ref{antennatemp}:

\begin{equation}
kT_{A}=\frac{1}{2}\eta_{R}\eta_{A}S_{\nu}A
\label{antennatemp}
\end{equation}

where $k$ is the Boltzmann constant, $\eta_{A}$ is the aperture efficiency (0.3 for these observations), $\eta_{R}$ is the resolution correction and A  is the effective area of the antenna. This equation yields a value $T_{\rm A}$ = 6.8K. Correcting this value for sideband gains, atmospheric emission and main beam efficiency gives $T_{\rm mb}$ = 14.4 K. This means that the line to continuum ratio at 500 GHz is $3.7\div14.4$ = 0.25. This lies between the value of 10\% found by Greaves $\&$ White (\cite{greaves:white}) and 30-40\% found by Sutton \al (\cite{sutton:etal}) at lower frequencies. We stress that this value is the best estimate lower limit to the line-to-continuum ratio that can be made with the present data - a further measurement that simultaneously measures the continuum and line flux is desirable to obtain a more accurate value - and would for example be valuable when trying to understand the relative contributions of thermal and line emission observed at far infrared wavelengths by bolometer detectors, and far-IR fourier transform spectrometers. The value we have estimated is consistent with the prediction given by Groesbeck (1995 - also reported in Schilke \al 2001) that the contribution of line emission to the total flux of Orion should drop from $\sim$ 50$\%$ at 800 \mum to 10$\%$ at 450 \mum.

\section{Conclusions}
A spectral line survey of the hot core region of the OMC-1 cloud core was obtained over the frequency intervals 455 - 468 GHz and from 491.8 - 507.4 GHz.

\begin{itemize}
\item
In this spectral line survey we find a total of 254 lines toward the hot core position at the centre of the Orion Nebula.
\item
Spectral lines of SO, SO$_{\rm 2}$ and CH$_3$OH and from large organic molecules such as (CH$_{\rm 3}$)$_2$O, CH$_{\rm 3}$CN, C$_{\rm 2}$H$_{\rm 5}$CN, HCOOCH$_{\rm 3}$ and C$_{\rm 2}$H$_{\rm 3}$CN make up $\sim$ 72\% of the total number of lines
\item
A total of 33 lines (13\% of the total lines identified) could not be associated with
known molecular transitions and are designated U-lines - this is comparable to the percentages of U-lines seen in other spectral lines surveys of Orion.
\item
The rotational temperatures and column densities derived using standard rotation diagram analysis techniques, showed ranges from 70 - 300 K, and 10$^{14}$ -- 10$^{17}$ cm$^2$ respectively (we exclude the CH$_3$OH line from this, as the error is large).

\item
The diameter of the region responsible for emitting the C$_{\rm 2}$H$_{\rm 5}$CN and C$_{\rm 2}$H$_{\rm 3}$CN lines is estimated to be $\sim$ 10\arcsec, based on a comparison with earlier data for the same lines.

\item
The lower limit to the line-to-continuum ratio from the hot core region is 0.25, which appears to be consistent with predictions based on modeling the expected intensities of lines from the JPL Molecular Spectroscopy database.

\end{itemize}

\section{Acknowledgments}

We acknowledge discussions with Prof S. Saito at Fukui University, Dr Y. Fukuyama at the Institute of Physical and Chemical Research, Prof John Pearson, for discussions about frequency assignments of molecules, and the referee for helpful comments.

\end{document}